\documentclass[pdflatex,aps,twocolumn,superscriptaddress,prd,fleqn,showpacs,nofootinbib,preprintnumbers]{revtex4-1}
\usepackage{amssymb,amsmath,epsfig,color}
\usepackage{times,txfonts}
\usepackage{float} 
\usepackage[caption=false]{subfig} 
\usepackage[nodisplayskipstretch]{setspace}
\usepackage[normalem]{ulem}

 \usepackage{ifpdf}
 \ifpdf
 \usepackage[pdftex]{hyperref}
 \else
 \usepackage[hypertex]{hyperref}
 \fi

 \hypersetup{
   pdftitle={},%
   pdfauthor={},%
   pdfsubject={},%
   pdfkeywords={},%
   pdfstartview={},%
   bookmarksopen=true, breaklinks=true, debug=true, %
   colorlinks=true, linkcolor=red, citecolor=blue, urlcolor=blue
 }

\newcommand{\sumint}{\sum_{X}\hspace{-0.5cm}\int}
\newcommand{\sumintXY}{\sum_{\!\!\!X,Y}\hspace{-0.45cm}\int}
\newcommand{\sumintY}{\sum_{Y}\hspace{-0.5cm}\int}
\newcommand{\sumintYone}{\sum_{Y_1}\hspace{-0.55cm}\int}
\newcommand{\sumintYN}{\sum_{Y_N}\hspace{-0.65cm}\int}
\newcommand{\sumintYNone}{\sum_{Y_{N-1}}\hspace{-0.55cm}\int}
\newcommand{\sumintYNonein}{\sum_{Y_{N-1}}\hspace{-0.85cm}\int}
\newcommand{\sumintYp}{\sum_{Y^\prime}\hspace{-0.55cm}\int}

\def\to{\rightarrow}

\def\beq{\begin{equation}}
\def\eeq{\end{equation}}
\def\beeq{\begin{eqnarray}}
\def\eeeq{\end{eqnarray}}
\def\beal{\begin{align}}
\def\eeal{\end{align}}

\def\nn{\nonumber}
\def\b0{b_0}

\def\ID{1 \kern -.45 em 1}

\def\slash#1{\setbox0=\hbox{$#1$}               
        \dimen0=\wd0                            
        \setbox1=\hbox{/} \dimen1=\wd1          
        \ifdim\dimen0>\dimen1                   
        \rlap{\hbox to \dimen0{\hfil/\hfil}}    
        #1                                      
        \else              
        \rlap{\hbox to \dimen1{\hfil$#1$\hfil}} 
        /                                       
        \fi}                                    %

\allowdisplaybreaks

\newcommand{\bra}[1]{\langle #1 |}
\newcommand{\ket}[1]{| #1 \rangle}
\newcommand{\vb}{\biggl|}

\begin{document}

\title{
Operator Constraints for Twist-3 Functions and Lorentz Invariance Properties of Twist-3 Observables}

\author{Koichi Kanazawa}
\email{tuf29138@temple.edu}
\affiliation{Department of Physics, SERC, Temple University, Philadelphia, PA 19122, USA}

\author{Yuji Koike}
\email{koike@phys.sc.niigata-u.ac.jp}
\affiliation{Department of Physics, Niigata University, Ikarashi, Niigata 950-2181, Japan}

\author{Andreas Metz}
\email{metza@temple.edu}
\affiliation{Department of Physics, SERC, Temple University, Philadelphia, PA 19122, USA}

\author{Daniel Pitonyak}
\email{dpitonyak@quark.phy.bnl.gov}
\affiliation{RIKEN BNL Research Center, Brookhaven National Lab, Upton, New York 11973, USA}

\author{Marc Schlegel}
\email{marc.schlegel@uni-tuebingen.de}
\affiliation{Institute for Theoretical Physics, University of T\"{u}bingen, Auf der Morgenstelle 14, D-72076 T\"{u}bingen, Germany}


\begin{abstract}
We investigate the behavior under Lorentz tranformations of perturbative coefficient functions in a collinear twist-3 formalism relevant for high-energy observables including transverse polarization of hadrons. We argue that those perturbative coefficient functions can, {\it a priori}, acquire quite different yet Lorentz-invariant forms in various frames. This somewhat surprising difference can be traced back to a general dependence of the perturbative coefficient functions on lightcone vectors which are introduced by the twist-3 factorization formulae and which are frame-dependent.  One can remove this spurious frame dependence by invoking so-called Lorentz invariance relations (LIRs) between twist-3 parton correlation functions. Some of those relations for twist-3 distribution functions were discussed in the literature before. In this paper we derive the corresponding LIRs for twist-3 fragmentation functions. We explicitly demonstrate that these LIRs remove the lightcone vector dependence by considering transverse spin observables in the single-inclusive production of hadrons in lepton-nucleon collisions, $\ell N\to hX$. With the LIRs in hand, we also show that twist-3 observables in general can be written solely in terms of three-parton correlation functions.
\end{abstract}

\date{\today}

\maketitle


\section{Introduction}

Observables in single-inclusive processes that arise from the transverse spin of hadrons have been studied for close to 40 years.  The first measurements of so-called single-spin asymmetries (SSAs) were made at Argonne National Lab~\cite{Klem:1976ui} and FermiLab~\cite{Bunce:1976yb} in the mid-1970s, and then continued with FermiLab again in the 1990s~\cite{Adams:1991rw}, and now most recently at AGS~\cite{Krueger:1998hz,Allgower:2002qi} and RHIC~\cite{Adams:2003fx,Adler:2005in,Lee:2007zzh,:2008mi,Adamczyk:2012qj,Adamczyk:2012xd,Bland:2013pkt,Adare:2013ekj} at Brookhaven National Lab.  All of these experiments involved proton-(anti)proton collisions that produced a single hadron (or jet), and only one of the hadrons involved in the process was transversely polarized.  These reactions arise as twist-3 observables that probe quark-gluon-quark (or tri-gluon) correlations in hadrons~\cite{Efremov:1981sh,Qiu:1991pp,Qiu:1991wg,Qiu:1998ia}.  In addition, there are also twist-3 double-spin asymmetries (DSAs) that arise when two hadrons are polarized:~one in the longitudinal direction and one in the transverse.  These so far have not been measured at proton-(anti)proton colliders, but the PHENIX Collaboration at RHIC has proposed to run such an experiment~\cite{PHENIX:BeamUse}.

Much theoretical work has been performed over the years to develop the collinear twist-3 formalism needed to describe these proton-proton 
SSAs~\cite{Efremov:1981sh,Qiu:1991pp,Qiu:1991wg,Qiu:1998ia,Kanazawa:2000hz,Eguchi:2006qz,Kouvaris:2006zy,Eguchi:2006mc,Koike:2006qv,Koike:2007rq,Koike:2009ge,Yuan:2009dw,Kang:2010zzb,Kanazawa:2010au,Metz:2012ct,Kanazawa:2013uia,Beppu:2013uda} and DSAs~\cite{Koike:2008du,Metz:2010xs,Lu:2011th,Liang:2012rb,Metz:2012fq,Hatta:2013wsa,Koike:2015yza}, including recent extensions to include fragmentation effects~\cite{Yuan:2009dw,Kang:2010zzb,Metz:2012ct,Kanazawa:2013uia,Koike:2015yza} and tri-gluon correlations~\cite{Beppu:2013uda,Hatta:2013wsa,Beppu:2010qn,Koike:2011ns,Koike:2011mb,Koike:2011nx}.  Nevertheless, whether this framework can provide a definitive explanation of these observables (especially SSAs) remains an open question.  For example, within collinear twist-3 factorization, SSAs in $p^\uparrow p\to h\, X$ at first were thought to arise from quark-gluon-quark correlations inside the transversely polarized proton through the so-called Qiu-Sterman (QS) function~\cite{Qiu:1991pp,Qiu:1991wg,Qiu:1998ia,Kouvaris:2006zy}.  However, this description turned out to be inconsistent with the Sivers effect~\cite{Sivers:1989cc}, a closely related observable that arises in semi-inclusive deep-inelastic scattering (SIDIS), and became known as the ``sign-mismatch" problem~\cite{Kang:2011hk}.  Later work actually gave strong evidence that the QS function could not be the main source of proton-proton SSAs~\cite{Kang:2012xf,Metz:2012ui}.  This led to a recent analysis that indicated these asymmetries could very well come from quark-gluon-quark fragmentation functions~\cite{Kanazawa:2014dca}, but more work is needed in order to ultimately resolve what causes them. (A frequently used alternative approach is the so-called Generalized Parton Model~\cite{Anselmino:2012rq,Anselmino:2013rya,Anselmino:2014eza}.) 

Because of the complicated nature of proton-proton SSAs, and the difficulty so far in isolating their source, one may instead hope to gain insight into their origin by looking at SSAs (and DSAs) in theoretically simpler lepton-nucleon scattering reactions. The prototype of such a reaction is the deep-inelastic scattering (DIS) process $\ell\,N\to \ell\,X$ where the scattered lepton is detected in the final state. Much of the currently available information about the partonic substructure of the nucleon originates from this process. Spin effects in DIS are conventionally described in terms of the $g_1$ and $g_2$ structure functions which can be probed in the DSA of a longitudinally polarized lepton and a longitudinally ($g_1$) or transversely ($g_2$) polarized nucleon. The latter observable can be analyzed in the collinear twist-3 formalism as well. As is well-known, the $g_2$ structure function receives a major contribution from the leading twist part of the $g_1$ structure function, which is typically referred to as the Wandzura-Wilczek contribution~\cite{Wandzura:1977qf}. Interestingly, deviations of the Wandzura-Wilczek approximation from experimental data can be explained by the aforementioned quark-gluon-quark correlations~\cite{Accardi:2009au}.

On the other hand, a transverse nucleon SSA in DIS can only be generated through higher order QED effects~\cite{Metz:2012ui,Christ:1966zz,TPE1,TPE2,TPE5}. Although one may in principle get access not only to quark-gluon-quark correlations~\cite{TPE5} but also to quark-photon-quark correlations~\cite{Metz:2012ui} in the nucleon by measuring this SSA, an experimental determination is difficult because of the small size of the observable. Nonetheless, recent measurements have been performed by HERMES~\cite{TPEHERMES} on a polarized proton target and at Jefferson Lab~\cite{TPEJLab} on a polarized $\mathrm{He^3}$ target, where the first non-vanishing effect has been seen. 

One may consider the process of single-hadron production from lepton-nucleon collisions, i.e., $\ell\,N\to h\,X$, as a hybrid of single-hadron production from $pp$-collisions and DIS. In the case of the former, the replacement of the unpolarized proton in $p^\uparrow p \to h\,X$ by an unpolarized lepton leads to tremendous theoretical simplifications. This is because many twist-3 functions, such as tri-gluon correlations or chiral-odd quark-gluon-quark correlations (in the unpolarized proton), which compete in a leading-order (LO) description for $pp$-collisions, will not appear in the collinear twist-3 framework for $\ell\,N\to h\,X$ (at LO).  Also, one must calculate significantly less Feynman diagrams to determine the perturbative coefficients in $\ell N$-collisions. 

Moreover, there are other motivations behind analyzing $\ell\,N\to h\,X$.  First, one can scrutinize the theoretical techniques behind collinear twist-3 calculations through checks of gauge invariance and Lorentz invariance (frame-independence).  Second, data exists for SSAs ($\ell \,p^\uparrow \to \{\pi,\,K\}\,X$~\cite{Airapetian:2013bim}, $\ell \,n^\uparrow \to \{\pi,\,K\}\,X$~\cite{Allada:2013nsw}, $\ell \,p\to \Lambda^\uparrow X$~\cite{Airapetian:2014tyc}) and DSAs ($\vec{\ell}\,n^\uparrow\to \{\pi,\,K\}\, X$~\cite{Zhao:2015wva}) in this reaction, and already several theoretical works in the literature have analyzed these observables~\cite{Kang:2011jw,Anselmino:2014eza,Gamberg:2014eia,Kanazawa:2014tda,Kanazawa:2015jxa}.  Lastly, one can attempt to push the twist-3 framework to next-to-leading-order for this process~\cite{Hinderer:2015hra,Schlegel:2015hga}.  

In this paper we address the first item, namely the gauge invariance and Lorentz invariance of twist-3 observables, especially the ones that involve twist-3 fragmentation functions.  These issues have only been studied in bits and pieces in the literature.  For example, only in Ref.~\cite{Kanazawa:2014tda} were both the gauge invariance and Lorentz invariance of the result checked, where for the latter a LIR was needed to demonstrate the frame-independence of the distribution term.  We trace the origin of why twist-3 observables are not manifestly frame independent (i.e., why one must employ LIRs) and derive such LIRs at the operator level for both twist-3 distribution and fragmentation functions.  We then compute several twist-3 transverse spin observables for $\ell\, N\to h\,X$ and explicitly demonstrate their Lorentz invariance by using the LIRs.  With the LIRs in hand, we also show how one can write twist-3 observables solely in terms of three-parton correlation functions.  

The paper is organized as follows:~in Sec.~\ref{s:der} we introduce the relevant twist-3 correlation functions and, in particular, derive the LIRs between them at the operator level.  We also show how all twist-3 functions can be written in terms of three-parton correlations.  Next, in Sec.~\ref{s:LI} we discuss in detail why twist-3 observables, {\it a priori}, could be frame-dependent and demonstrate (using asymmetries in various polarized $\ell\,N\to h\,X$ processes) how LIRs remove the frame dependence.  We write final results solely in terms of three-parton correlations in order to emphasize that these are the essential functions in the description of twist-3 SSAs and DSAs studied here.  Finally, in Sec.~\ref{s:concl} we summarize our work and give our conclusions.


\section{Derivation of Lorentz invariance relations between twist-3 correlators \label{s:der}}
In this section we derive the LIRs that involve twist-3 correlation functions, where we restrict ourselves to correlation functions of quark fields.  We will investigate twist-3 matrix elements for parton correlations in the nucleon and parton correlations in the fragmenting hadron. Although, in general, twist-3 matrix elements lack a probability interpretation (in contrast to ordinary twist-2 parton distribution functions (PDFs) or fragmentation functions (FFs)), for the sake of brevity we will nevertheless refer to twist-3 correlations in the nucleon as {\it parton distributions} and to twist-3 correlations in the fragmenting hadron as {\it fragmentation functions}.  After defining and parameterizing these correlators, we then review a class of previously known relations, called equation of motion (e.o.m.)~relations, before moving on to derive the aforementioned LIRs, where those for fragmentation functions are new from this work.

\subsection{Parameterizations of matrix elements of twist-3 operators \label{s:par}}

We present the common parameterizations of twist-3 matrix elements that are relevant for the pQCD description of transverse spin observables in inclusive high-energy processes.  Unlike the case of twist-2 reactions, there are several different matrix elements that can contribute to twist-3 observables. We will refer to them as {\it intrinsic}, {\it kinematical}, and {\it dynamical} correlation functions.  

We start with the {\it intrinsic} twist-3 functions.  These are given in terms of ordinary collinear two-quark operators, where, in the language of lightcone quantization, the so-called ``bad component''~\cite{Jaffe:1996zw} of one of the quark fields is probed. They are defined in terms of a distribution correlator $\Phi$,
\begin{eqnarray}
\Phi^q_{ij}(x) &=&\int_{-\infty}^{\infty}\frac{d\lambda}{2\pi}\,\mathrm{e}^{i\lambda x}\langle P,S|\,\bar{q}_j(0)\,\mathcal{W}[0\,;\,\lambda n]\, q_i(\lambda n)\,|P,S\rangle\,,\label{Phix}
\end{eqnarray}
and a fragmentation correlator $\Delta$,
\begin{eqnarray}
\Delta^{q}_{ij}(z) & = & \frac{1}{N_{c}}\sumint\int_{-\infty}^{\infty}\frac{d\lambda}{2\pi}\,\mathrm{e}^{-i\frac{\lambda}{z}}\langle0|\, \mathcal{W}[ \pm \infty m\,;\,0]\,q_i(0)\,|P_{h}S_{h};\,X\rangle \nonumber\\
&& \times \langle P_{h}S_{h};\,X|\,\bar{q}_j(\lambda m)\,\mathcal{W}[\lambda m\,;\,\pm\infty m]\,|0\rangle\,.\label{Deltaz}
\end{eqnarray}
where $q$ is a quark field and $\mathcal{W}$ is a Wilson line that renders these matrix elements color gauge invariant.  The number of colors is indicated by $N_c$. The two definitions (\ref{Phix}) and (\ref{Deltaz}) incorporate two lightcone vectors $n^\mu$ and $m^\mu$. The momenta $P^\mu$ and $P^\mu_h$ of the nucleon and the outgoing hadron, respectively, are themselves approximately lightcone vectors with $P^2\simeq 0$ and $P_h^2\simeq 0$ for negligible nucleon/hadron masses. Hence, the vector $n$ is defined to represent the ``adjoint" lightcone direction with $n^2=0$ and $P\cdot n=1$. Likewise, we have $m^2=0$ and $P_h\cdot m=1$. We note that those two conditions do \emph{not} completely fix the adjoint lightcone vectors $n^\mu$ and $m^\mu$ for given hadron momenta $P^\mu$ and $P_h^\mu$. In fact, there is a certain arbitrariness in the choice of $n^\mu$ and $m^\mu$ which will be discussed further in the following section. Only after the lightcone vectors $n$ and $m$ are chosen or fixed can one identify the ``transverse" directions with respect to $P$ and $n$ through $a_T^{\mu}\equiv g_T^{\mu \nu}a_{\nu}=(g^{\mu \nu}-P^{\mu}n^\nu-P^\nu n^\mu)a_\nu$ (labeled by a subscript $T$) as well as a ``transverse" direction with respect to $P_h$ and $m$ through $a_\perp^\nu\equiv g_\perp^{\mu \nu}a_\nu=(g^{\mu \nu}-P_h^\mu m^\nu-P_h^\nu m^\mu)a_\nu$ (labeled by a subscript $\perp$).
 The correlators $\Phi$ and $\Delta$ defined in (\ref{Phix}) and (\ref{Deltaz}) carry two Dirac indices $i,j$ and are therefore Dirac matrices. They can be parameterized in terms of scalar functions called distribution functions in the case of $\Phi$ and fragmentation functions in the case of $\Delta$. Those parameterizations were presented in the literature (cf.~\cite{Mulders:1995dh,Boer:1997mf,Goeke:2005hb}), and we just give the results for $\Phi$,
\begin{eqnarray}
\Phi^{q}(x) & \!\!\!=\!\!\! & \frac{1}{2}\Bigg(\slash P\, f_{1}^{q}(x)+M\, e^{q}(x)\nonumber \\
 &  & -M\,(S\cdot n)\,\slash P\gamma_{5}\, g_{1}^{q}(x)+\tfrac{1}{2}M^{2}\,(S\cdot n)\,[\slash P,\slash n]\gamma_{5}\, h_{L}^{q}(x)\nonumber \\
 &  & -\frac{1}{2}[\slash P, 
\slash  S]\gamma_{5}\, h_{1}^{q}(x)  -M\,(\slash S-(S\cdot n)\slash P)\gamma_{5}\, g_{T}^{q}(x)\Bigg)\,,\label{eq:PhixParam}
\end{eqnarray}
and $\Delta$,
\begin{eqnarray}
\Delta^{q}(z) &\!\!\! =\!\!\! & \,\frac{1}{z}\Bigg(\slash P_{h}\, D_{1}^{q}(z)+M_{h}\, E^{q}(z)+\tfrac{i}{2}M_{h}[\slash P_{h},\slash m]\, H^{q}(z)\nonumber \\
 &  & -M_{h}\,(S_{h}\cdot m)\,\slash P_{h}\gamma_{5}\, G_{1}^{q}(z)\nonumber\\
& & +\tfrac{1}{2}M_{h}^{2}(S_{h}\cdot m)\,[\slash P_{h},\slash
 m]\gamma_{5}\, H_{L}^{q}(z) - M_h^2 (S_h\cdot m) i \gamma_5 \,
 E_L^q(z)\nonumber \\
 &  & -\frac{1}{2}[\slash P_{h},
\slash S_{h}]\gamma_{5}\, H_{1}^{q}(z)-M_{h}\,\epsilon^{P_{h}m\alpha S_{h}}\gamma_{\alpha}\, D_{T}^{q}(z)\nonumber \\
 &  & -M_{h}\,(\slash S_{h}-(S_{h}\cdot m)\slash P_{h})\gamma_{5}\, G_{T}^{q}(z)\Bigg)\,.\label{eq:DeltazParam}
\end{eqnarray}
We encounter the spin vectors $S^\mu$ and $S_h^\mu$ for spin-1/2 hadrons in those parameterizations. They obey the conditions $S^2=-1$ and $P\cdot S=0$, as well as $S_h^2=-1$ and $P_h\cdot S_h=0$. In addition, the nucleon and hadron masses, $M$ and $M_h$, are introduced to match the mass dimensions.  Our convention for the Levi-Civita tensor is $\epsilon^{0123} = 1$.
The correlators $\Phi$ and $\Delta$ are parameterized up to twist-3 correlation functions.  Twist-4 functions are dropped as they do not serve the purpose of this paper. 

The parton distribution functions (PDFs) $f_1$, $g_1$, and $h_1$ are the leading twist unpolarized, helicity, and transversity parton densities that appear in (polarized) twist-2 observables~\cite{Ralston:1979ys,Jaffe:1991kp,Cortes:1991ja}. The fragmentation functions (FFs) $D_1$, $G_1$, and $H_1$ are the twist-2 counterparts on the fragmentation side for these PDFs. The remaining twist-3 correlation functions lack a probabilistic interpretation and are what we call {\it intrinsic} twist-3 functions.
Each function defined in (\ref{eq:PhixParam}) and (\ref{eq:DeltazParam}) has a support $-1<x<1$ or $0<z<1$.  
Antiquark distribution and fragmentation functions are defined by replacing the non-local operators in 
(\ref{Phix}) and (\ref{Deltaz}) by their charge conjugated operators, which results in $q\to C\bar{q}^T$ and $\bar{q}\to -q^T C^{-1}$,
where $C \equiv i\gamma^2 \gamma^0$ is the charge conjugation matrix.  
The antiquark distributions defined this way are related to the quark distributions as
$f^{\bar{q}}(x)=-f^q(-x)$ for $f=f_1,\ h_1,\ h_L$ and $f^{\bar{q}}(x)=f^q(-x)$ for $f=g_1,\ g_T,\ e$, 
while no such relations exist for the fragmentation functions.  

The second class of twist-3 matrix elements are the {\it kinematical} correlation functions. These types of functions originate from matrix elements that depend on the parton transverse momentum. We consider the following quantities,
\begin{eqnarray}
\Phi^q_{ij}(x,k_T)&\!\!\!=\!\!\!& \int_{-\infty}^{\infty}\frac{d\lambda}{2\pi}\int\frac{d^2z_T}{(2\pi)^2}\,\mathrm{e}^{ix\lambda+ik_T\cdot z_T}\times\nonumber\\
& &\hspace{-1cm} \langle P,S|\,\bar{q}_j(0)\,\mathcal{W}[0\,;\,\infty n]\,\times\,\nonumber\\
&& \hspace{-1cm}\mathcal{W}[\infty n\,;\,\infty n+\infty z_T]\,\times\,\mathcal{W}[\infty n+\infty z_T\,;\,\infty n+z_T]\,\times\,\nonumber\\
&& \hspace{-1cm}\mathcal{W}[\infty n+z_T\,;\,\lambda n+z_T]\,q_i(\lambda n+z_T)\,|P,S\rangle\,,\label{PhiTMD}
\end{eqnarray}
\begin{eqnarray}
\Delta^q_{ij}(z,p_\perp)& \!\!\!=\!\!\! & \frac{1}{N_{c}}\sumint\int_{-\infty}^{\infty}\frac{d\lambda}{2\pi}\int \frac{d^2 z_\perp}{(2\pi)^2}\,\mathrm{e}^{-i\frac{\lambda}{z}-ip_\perp\cdot z_\perp}\times\label{DeltaTMD}\\
& &\hspace{-1.8cm}\langle0|\mathcal{W}[\pm\infty m+\infty z_\perp\,;\,\pm \infty m]\times \mathcal{W}[ \pm \infty m\,;\,0]\,q_i(0)|P_{h}S_{h};\,X\rangle \times \nonumber\\
&& \hspace{-0.2cm} \langle P_{h}S_{h};\,X|\,\bar{q}_j(\lambda m+z_\perp)\,\times\nonumber\\
&&\hspace{-1.8cm}\mathcal{W}[\lambda m+z_\perp\,;\,\pm\infty m+z_\perp]\times \mathcal{W}[\pm \infty m+z_\perp\,;\,\pm\infty m+\infty z_\perp]\,|0\rangle\,.\nonumber
\end{eqnarray}
The kinematical twist-3 correlators $\Phi^{\rho}_\partial(x)$ and $\Delta^{\rho}_\partial(z)$ are then generated by weighting the correlators $\Phi(x,k_T)$ and $\Delta(z,p_\perp)$ by the parton transverse momentum and integrating over it,
\begin{eqnarray}
\Phi^{q,\rho}_{\partial,ij}(x)& = & \int d^2 k_T\, k_T^\rho\, \Phi^q_{ij}(x,k_T)\,,\label{partPhix}\\
\Delta^{q,\rho}_{\partial,ij}(z) & = & \int d^2p_\perp\,p^\rho_\perp\, \Delta^q_{ij}(z,p_\perp)\,.\label{partDeltaz} 
\end{eqnarray}
The parameterization of those matrix elements can again be found in Refs.~\cite{Mulders:1995dh,Boer:1997mf,Goeke:2005hb},
\begin{eqnarray}
\Phi_{\partial}^{q,\rho}(x) & \!\!\!=\!\!\! & \frac{1}{2}\Bigg(M\epsilon^{Pn\rho S}\slash P\, f_{1T}^{\perp(1),\, q}(x)\nonumber\\
&& -M(S^{\rho}-P^{\rho}(n\cdot S))\,\slash P\gamma_{5}\, g_{1T}^{(1),\, q}(x)\nonumber \\
 &  & +\frac{1}{2}M^{2}(n\cdot S)\,\left([\slash P,\gamma^{\rho}]\gamma_{5}-P^{\rho}[\slash P,\slash n]\gamma_{5}\right)\, h_{1L}^{\perp(1),\, q}(x)\nonumber \\
 &  & +\frac{i}{2}M\,\left([\slash P,\gamma^{\rho}]-P^{\rho}[\slash P,\slash n]\right)\, h_{1}^{\perp(1),\, q}(x)\Bigg)\,,\label{eq:ParamPhipart}
\end{eqnarray}
\begin{eqnarray}
\Delta_{\partial}^{q,\rho}(z) & \!\!\!=\!\!\! & \frac{1}{z}\Bigg(M_{h}\epsilon^{P_{h}m\rho S_{h}}\slash P_{h}\, D_{1T}^{\perp(1),\, q}(z)\nonumber \\
 &  &\hspace{-0.3cm} -M_{h}(S_{h}^{\rho}-P_{h}^{\rho}(m\cdot S_{h}))\,\slash P_{h}\gamma_{5}\, G_{1T}^{\perp(1),\, q}(z)\nonumber \\
 &  &\hspace{-0.3cm} +\frac{1}{2}M_{h}^{2}(m\cdot S_{h})\,\left([\slash P_{h},\gamma^{\rho}]\gamma_{5}-P_{h}^{\rho}[\slash P_{h},\slash m]\gamma_{5}\right)\, H_{1L}^{\perp(1),\, q}(z)\nonumber \\
 &  &\hspace{-0.3cm} +\frac{i}{2}M_{h}\,\left([\slash P_{h},\gamma^{\rho}]-P_{h}^{\rho}[\slash P_{h},\slash m]\right)\, H_{1}^{\perp(1),\, q}(z)\Bigg)\,.\label{eq:ParamDeltapart}
\end{eqnarray}
As our notation indicates, the kinematical twist-3 functions can be written as transverse-momentum-moments of transverse momentum dependent (TMD) parton distribution and fragmentation functions.  These moments are defined as 
\begin{align}
f^{(1)}(x) &= \int\!d^2k_T\frac{\vec{k}_T^{\,2}} {2M^2}\,f(x,k_T^2)\,,\\
D^{(1)}(z) &=z^2\int\!d^2p_\perp\frac{\vec{p}_\perp^{\,2}} {2M_h^2}\,D(z,z^2p_\perp^2)\,.
\end{align}
Antiquark distribution and fragmentation functions can also be defined  for the kinematical twist-3 functions 
as in the case of the functions in  (\ref{eq:PhixParam}) and (\ref{eq:DeltazParam}). 
For the distribution functions, quark and antiquark ones are related as 
$f^{\bar{q}}(x) = f^q(-x)$ for $f=f_{1T}^{\perp (1)},\ h_{1L}^{\perp (1)},\ h_1^{\perp (1)}$
and  $f^{\bar{q}}(x) = -f^q(-x)$ for $f=g_{1T}^{\perp (1)}$.  No such relations exist for the fragmentation functions.

The third type of twist-3 correlators can be generated from matrix elements that involve three parton fields.  These consist typically of two quark fields $q$ and one gluonic field-strength tensor $F^{\mu \nu}$. Such terms are called {\it dynamical} or {\it genuine} or {\it pure} twist-3 correlation functions in the literature. Here we will refer to them as {\it dynamical} twist-3 matrix elements. They exist for both distribution and fragmentation, and are defined as follows: 
\begin{eqnarray}
\Phi^{q,\rho}_{F,ij}(x,x^\prime) & \!\!\!=\!\!\! & \int_{-\infty}^\infty\frac{d\lambda}{2\pi}\int_{-\infty}^\infty\frac{d\mu}{2\pi}\,\mathrm{e}^{ix^\prime \lambda+i(x-x^\prime)\mu}\times\label{PhiFxx'}\\
&&\hspace{-1.6cm}\langle P,S|\,\bar{q}_j(0)\,\mathcal{W}[0\,;\,\mu n]\,ign_{\eta}F^{\eta \rho}(\mu n)\,\mathcal{W}[\mu n\,;\,\lambda n]\,q_i(\lambda n)\,|P,S\rangle ,\nonumber\\[0.3cm]
\Delta^{q,\rho}_{F,ij}(z,z^\prime)& \!\!\!=\!\!\! & \frac{1}{N_c}\sumint  \int_{-\infty}^\infty\frac{d\lambda}{2\pi}\int_{-\infty}^\infty\frac{d\mu}{2\pi}\,\mathrm{e}^{i\frac{\lambda}{z^\prime}+i(\frac{1}{z}-\frac{1}{z^\prime})\mu} \times\nonumber\\
&& \hspace{-0cm}\langle 0|\,\mathcal{W}[\pm \infty m\,;\,\mu m]\,igm_\eta F^{\eta \rho}(\mu m) \times \nonumber\\
&& \mathcal{W}[\mu m\,;\, \lambda m] q_i(\lambda m)\,|P_hS_h\,;\,X\rangle \times \nonumber\\
&&\hspace{-0cm}\langle P_hS_h\,;\,X|\,\bar{q}_j(0)\,\mathcal{W}[0\,;\, \pm\infty m]\,|0\rangle\,.\label{DeltaFzz'}
\end{eqnarray}
Since three parton fields are involved in these matrix elements, the correlators $\Phi^{\rho}_F$ and $\Delta^{\rho}_F$ depend on two partonic momentum fractions $x$ and $x^\prime$ or $z$ and $z^\prime$. As before, the matrix elements can be parameterized in terms of scalar functions~(see, e.g.,~\cite{Efremov:1981sh,Qiu:1991wg,Eguchi:2006qz,Meissner:2009,Zhou:2009jm,Metz:2012ct,Kanazawa:2013uia,Kanazawa:2015jxa}), and they read
\begin{eqnarray}
\Phi_{F}^{q,\rho}(x,x^{\prime}) & = & \frac{M}{2}\Bigg(\epsilon^{Pn\rho S}\,\slash P\, iF_{FT}^{q}(x,x^{\prime})\label{eq:ParamPhiF}\\
&&-(S^{\rho}-P^{\rho}(n\cdot S))\,\slash P\gamma_{5}\, G_{FT}^{q}(x,x^{\prime})\nonumber \\
 &  & +\frac{i}{2}\left([\slash P,\gamma^{\rho}]-P^{\rho}[\slash P,\slash n]\right)\, iH_{FU}^{q}(x,x^{\prime})\nonumber \\
 &  & \hspace{-1.2cm}+\frac{M}{2}(n\cdot S)\,\left([\slash P,\gamma^{\rho}]\gamma_{5}-P^{\rho}[\slash P,\slash n]\gamma_{5}\right)\, H_{FL}^{q}(x,x^{\prime})\Bigg)\,,\nonumber
\end{eqnarray}
\begin{eqnarray}
\Delta_{F}^{q,\rho}(z,z^{\prime}) & = & \frac{M_{h}}{z}\Bigg(\epsilon^{P_{h}m\rho S_{h}}\slash P_{h}\,i\hat{D}_{FT}^{q}(z,z^{\prime})\label{eq:ParamDeltaF}\\
&&-(S_{h}^{\rho}-P_{h}^{\rho}(m\cdot S_{h}))\,\slash P_h \gamma_5\,\hat{G}_{FT}^{q}(z,z^{\prime})\nonumber \\
 &  & +\frac{i}{2}([\slash P_{h},\gamma^{\rho}]-P_{h}^{\rho}[\slash P_{h},\slash m])\,i\hat{H}_{FU}^{q}(z,z^{\prime})\nonumber \\
 &  & \hspace{-1.2cm}+\frac{M_h}{2}(m\cdot S_{h})\,([\slash P_{h},\gamma^{\rho}]\gamma_{5}-P_{h}^{\rho}[\slash P_{h},\slash m]\gamma_{5})\,\hat{H}_{FL}^{q}(z,z^{\prime})\Bigg)\,.\nonumber
\end{eqnarray}
We note that the twist-3 three-parton distributions are real-valued functions whereas
the twist-3 three-parton fragmentation functions are complex. Other parameterizations for $\Delta^{\rho}_F$ with different prefactors of $z$ may be found in the literature. We also note that $F_{FT}(x,x^{\prime})=F_{FT}(x^{\prime},x)$,
$H_{FU}(x,x^{\prime})=H_{FU}(x^{\prime},x)$, $G_{FT}(x,x^{\prime})=-G_{FT}(x^{\prime},x)$
and $H_{FL}(x,x^{\prime})=-H_{FL}(x^{\prime},x)$. Hence, $G_{FT}(x,x)=H_{FL}(x,x)=0$.
Each dynamical twist-3 distribution has support on $|x\,| \le 1$, $|x^{\prime}|\le1$, and $|x-x'|\le1$.  On the fragmentation
side we have in general $\Delta_{F}^{\rho}(z,z)=0$ as well as $\Delta_{F}^{\rho}(z,0)=0$~\cite{Meissner:2008yf}. We further prove in the Appendix that $\frac{\partial}{\partial (1/z^\prime)}\Delta_{F}^{\rho}(z,z^\prime)\Big|_{z^\prime = z}=0$. This identity is new and to the best of our knowledge has never been mentioned in the literature.
The support properties for dynamical twist-3 fragmentation functions are $0\le z\le1$ and $z<z^{\prime}<\infty$.

For each dynamical twist-3 function, a ``charge-conjugated" function can be defined by
considering the matrix element of the charge-conjugated non-local operator.
This results in the change of $q\to C\bar{q}^T$, $\bar{q}\to -q^T C^{-1}$ and $F_{\mu\nu} \to -F_{\mu\nu}^T$ in 
(\ref{eq:ParamPhiF}) and (\ref{eq:ParamDeltaF}).  
For each distribution function $F^{q}(x_1,x_2)$ ($F=F_{FT},\ G_{FT},\ H_{FU},\ H_{FL}$), we call the resulting distribution 
$F^{\bar{q}}(x_1,x_2)$. 
We then have the relations for the distribution functions as
$F^{\bar{q}}(x_1,x_2)=F^q(-x_2,-x_1)$ for $F=F_{FT},\ H_{FU},\ H_{FL}$ and $F^{\bar{q}}(x_1,x_2)=-F^q(-x_2,-x_1)$ for $F=G_{FT}$.  
For the fragmentation functions there is no such relations between the original function and the charge-conjugated one. 
One needs to keep in mind the support properties discussed above
when deriving relations among the twist-3 distribution and fragmentation functions.  

\subsection{QCD equation of motion relations}
In a next step we list relations between intrinsic, kinematical, and dynamical correlation 
functions that can be derived by applying the QCD e.o.m.,~$(\slash D (-y) +im_q)q(-y)=0$ and $\bar{q}(y)\left(\overleftarrow{\slash D}(y) - im_q \right)=0$ (where $D^\mu = \partial^\mu - igA^\mu$),~to the field 
operators in the matrix elements. Such relations were already mentioned in the literature, e.g., 
Refs.~\cite{Efremov:1981sh,Boer:1997bw,Bacchetta:2006tn,Eguchi:2006qz,Meissner:2009,Zhou:2009jm,Metz:2012ct,Kanazawa:2013uia,Kanazawa:2015jxa}.
The most prominent of these QCD e.o.m.~relations was extensively used in the theoretical description of the $g_2$ structure function in DIS. 
It connects the kinematical twist-3 function $g_{1T}^{(1)}$, the intrinsic twist-3 function $g_T$, and 
the dynamical twist-3 functions $F_{FT}$ and $G_{FT}$, as well as quark mass $m_q$ effects proportional to the transversity $h_1$, in the following way:
\begin{equation}
g_{1T}^{(1),q}(x)=x\,g_T^q(x)-\tfrac{m_q}{M}\,h_1^q(x)+
\mathcal{P}\!\!\int_{-1}^1\!\!dx^\prime \,\frac{F_{FT}^q(x,x^\prime)-G_{FT}^q(x,x^\prime)}{x-x^\prime},\label{EoMgT}
\end{equation}
where $\mathcal{P}$ denotes the principal value prescription. There are two more relations that can be derived from the QCD e.o.m.,
\begin{equation}
h_{1L}^{\perp (1),q}(x)= -\frac{x} {2}\,h_L^q(x)+\tfrac{m_q}{2M}\,g_1^q(x)-\,\mathcal{P}\int_{-1}^1dx^\prime \,\frac{H_{FL}^q(x,x^\prime)}{x-x^\prime},\label{EoMhL}
\end{equation}
and
\begin{equation}
x\,e^q(x)=\tfrac{m_q}{M}\,f_1^q(x)-2\,\mathcal{P}\int_{-1}^1dx^\prime\,\frac{H_{FU}^q(x,x^\prime)}{x-x^\prime}.\label{EoMe}
\end{equation}

Analogously, one can derive relations for twist-3 fragmentation functions from the QCD e.o.m. Since there is a richer structure in the parameterizations (\ref{eq:DeltazParam}), (\ref{eq:ParamDeltapart}), and (\ref{eq:ParamDeltaF}) due to the absence of a time-reversal constraint, one can derive more relations compared to the case of twist-3 distributions. The most relevant e.o.m.~relations are for unpolarized outgoing hadrons, 
\begin{equation}
H_1^{\perp (1),q}(z)=-\frac{H^{q}(z)}{2z}+\int_{z}^\infty\frac{dz^\prime}{z^{\prime 2}}\,
\frac{\Im[\hat{H}_{FU}^{q}(z,z^\prime)]}{\frac{1}{z}-\frac{1}{z^\prime}},\label{EoMH}
\end{equation} 
and
\begin{equation}
-\frac{E^{q}(z)}{2z}=\int_z^\infty \frac{dz^\prime}{z^{\prime 2}}\,
\frac{\Re[\hat{H}_{FU}^{q}(z,z^\prime)]}{\frac{1}{z}-\frac{1}{z^\prime}}-\frac{m_q} {2M_h}\,D_1^{q}(z)\,.\label{EoME}
\end{equation}
For a transversely polarized hadron one can derive the following relations between twist-3 fragmentation functions,
\begin{eqnarray}
D_{1T}^{\perp (1),q}(z)&=&-\frac{D_T^{q}(z)}{z}\nonumber\\
&&+\int_z^\infty \frac{dz^\prime}{z^{\prime 2}}\,\frac{\Im[\hat{D}_{FT}^{q}(z,z^\prime)]-\Im[\hat{G}_{FT}^{q}(z,z^\prime)]}{\frac{1}{z}-\frac{1}{z^\prime}},
\label{EoMDT}\\[0.5cm]
G_{1T}^{(1),q}(z)&=& \frac{G_T^{q}(z)}{z}-\frac{m_q} {M_h}\,H_1^{q} (z)\nonumber\\
&&\hspace{-0.3cm}+\int_z^\infty 
\frac{dz^\prime}{z^{\prime 2}}\,\frac{\Re[\hat{D}_{FT}^{q}(z,z^\prime)]-\Re[\hat{G}_{FT}^{q}(z,z^\prime)]}{\frac{1}{z}-\frac{1}{z^\prime}}\,.\label{EoMGT}
\end{eqnarray}
For completeness we also note that there are relations between twist-3
fragmentation functions for a longitudinally polarized hadron, and they read
\begin{eqnarray}
  \frac{E_L^{q}(z)}{2z} &=& - \int_{z}^{\infty} \frac{dz'}{z'^2} 
  \frac{\Im [\hat{H}_{FL}^{q}(z,z')]}{\frac{1}{z}-\frac{1}{z'}}, \\[0.5cm]
 H_{1L}^{\perp(1),q}(z) &=& -\frac{H_L^{q}(z)}{2z} 
  + \frac{m_q}{2 M_h} \, G_1^{q}(z)  - \int_{z}^{\infty} \frac{dz'}{z'^2} \frac{\Re [ \hat{H}_{FL}^{q}(z,z')
  ]}{\frac{1}{z}-\frac{1}{z'}}.\nn\\
\end{eqnarray}

\subsection{Lorentz invariance relations}

In this subsection we derive another set of relations between intrinsic,
kinematical, and dynamical twist-3 functions.  
Such relations can be obtained
from identities among the non-local
operators in which the fields are not necessarily located on the 
lightcone~\cite{Balitsky:1987bk,Balitsky:1989ry,Balitsky:1990ck}
and fully incorporate the constraints from Lorentz invariance.  
Combination of these resulting relations and the e.o.m.~relations discussed in the previous subsection
leads to the LIRs.  
The method employed here has been widely used to obtain the relations
among twist-3 distributions\,\cite{Eguchi:2006qz,Kodaira:1998jn,Balitsky:1987bk} and 
among twist-3 distribution amplitudes~\cite{Balitsky:1989ry,Braun:1989iv,Ball:1998sk}
relevant to exclusive processes.
Here we apply this method systematically to the case of twist-3 fragmentation
functions and derive the corresponding LIRs. 
As we will see in later sections, these LIRs
play a critical role to guarantee the frame independence of
twist-3 cross sections. 

\subsubsection{LIRs for twist-3 distribution functions}

Here we first present a brief review of the method to derive LIRs among the twist-3 distributions.  
A crucial ingredient is the following identity for the gauge invariant non-local
operator of the form~\cite{Balitsky:1987bk,Balitsky:1989ry}
\begin{eqnarray}
\hat{J}^{[\Gamma]}_{\rho\alpha}(y)\!\!&\equiv&\!\!y_\rho
  {\partial \over \partial y^\alpha} \bar{q}(y){\cal W}[y;-y]\Gamma q(-y)\nonumber\\
&& \hspace{-1cm}= y_\rho \, \bar{q}(y) \left( \overleftarrow{D}_\alpha (y) {\cal
 W}[y;-y] - {\cal W}[y;-y] {D}_\alpha (-y)\right) \Gamma q(-y)  \nn\\
&&\hspace{-1.3cm}\quad +\, i \, y_\rho \int_{-1}^1\,dt\,t\,\bar{q}(y) {\cal W}[y;ty] g F_{\alpha y}(ty)  {\cal W}[ty;-y] \Gamma q(-y)\,,
\label{nonlocal_dist}
\end{eqnarray}
where $y$ is a general space-time coordinate not necessarily on the lightcone ($y^2\neq 0$)
and $\Gamma$ is a Dirac matrix.  To get a relation among the distributions, we introduce
the derivative over the total translation as
\begin{eqnarray}
&&\partial_\rho\left\{ \bar{q}(y){\cal W}[y;-y]\Gamma' q(-y) \right\} \nonumber\\
&&\hspace{-0.4cm}\equiv \lim_{\xi^\rho\to 0} {d\over d\xi^\rho} \bar{q}(y+\xi){\cal W}[y+\xi;-y+\xi]\Gamma' q(-y+\xi)\nonumber\\
&& \hspace{-0.4cm}= \bar{q}(y)\overleftarrow{D}_\rho (y) {\cal W}[y;-y]\Gamma' q(-y)
+ \bar{q}(y)  {\cal W}[y;-y]\Gamma' {D}_\rho (-y)q(-y)\nonumber\\
&&\hspace{-0.4cm}\qquad + \,i \int_{-1}^1\,dt\,\bar{q}(y) {\cal W}[y;ty] gF_{\rho y}(ty)  {\cal W}[ty;-y] \Gamma' q(-y)\,,
\label{total_dist}
\end{eqnarray}
where $\Gamma'$ is another Dirac matrix. 
For simplicity in notation, we shall omit the gauge-links ${\cal W}[y;-y]$, ${\cal W}[y;ty]$, ${\cal W}[ty;-y]$, etc.~in the non-local operators below.  
Due to translational invariance, the forward matrix element of this total derivative vanishes:
\begin{eqnarray}
\bra{P,S} \partial_\rho\left\{ \bar{q}(y) 
\Gamma' q(-y) \right\} \ket{P,S}=0\,.
\label{total_dist0}
\end{eqnarray}
To obtain the relations among the twist-3 distributions, 
one needs to take the nucleon matrix element of (\ref{nonlocal_dist}) and the lightcone limit $y^2\to 0$.
A judicious choice of $\Gamma'$ in (\ref{total_dist}) and (\ref{total_dist0}), 
together with the QCD e.o.m.,
enables one to replace in this matrix element the terms containing the covariant derivative
by those containing the field strength tensor.  

For the relation among $g_T$, $g_1$, $F_{FT}$,  
and $G_{FT}$,
we consider the matrix element of (\ref{nonlocal_dist}) 
with $\Gamma = (g^{\rho\alpha} g_{\,\,\,\sigma}^\lambda - g_{\,\,\,\sigma}^\alpha
g^{\rho\lambda}) \gamma_\lambda \gamma_5$ 
and that of (\ref{total_dist0}) with $\Gamma'=y_\beta(\sigma^{\sigma
\beta}\gamma^\rho + \gamma^\rho \sigma^{\sigma \beta}) \gamma_5$.  
After some algebra,
one obtains the identity as
\begin{eqnarray}
&&\bra{P,S} y^\mu\left( {\partial \over \partial y^\mu} \bar{q}(y) \gamma_\sigma \gamma_5 q(-y)
- {\partial \over \partial y^\sigma} \bar{q}(y) \gamma_\mu \gamma_5 q(-y) \right)\ket{P,S}\nonumber\\
&&= \bra{P,S}2 m_q  \bar{q}(y) \sigma_{y \sigma} \gamma_5 q(-y)\ket{P,S}\nonumber\\
&&- \bra{P,S}\int_{-1}^1 dt\,t\, \bar{q}(y) g\,F_{\sigma y}(ty) \slash y\, i\,\gamma_5 q(-y)\ket{P,S}\nonumber\\
&&+ \bra{P,S}\int_{-1}^1 dt\,\bar{q}(y) g\,F^{\rho y}(ty) \epsilon_{\lambda y \sigma \rho}
\gamma^\lambda q(-y)\ket{P,S}, 
\end{eqnarray}
where we have already used the relation (\ref{total_dist0}) and the QCD e.o.m.   
Taking the lightcone limit ($y^2\to 0$) for this expression, one arrives at the following ``$\hat{J}$-operator relation''~\cite{Eguchi:2006qz,Kodaira:1998jn,Balitsky:1987bk}
\begin{eqnarray}
&& g_1^q(x) + x \frac{dg_T^q(x)}{dx} = {m_q\over M}{dh_1^q(x)\over dx}- {\cal P}\int_{-1}^1 dx' \frac{1}{x-x'} \nn\\
&& \times \left[ \left( \frac{\partial}{\partial x} + \frac{\partial}{\partial
	  x'} \right) F_{FT}^q(x,x') - 
  \left( \frac{\partial}{\partial x} - \frac{\partial}{\partial
   x'} \right) G_{FT}^q(x,x') \right].
\label{g1gT}
\end{eqnarray}
This equation can be solved to give $g_T$ in terms of
$g_1$, $F_{FT}$, and $G_{FT}$.  We shall postpone the solution to the next subsection.
Instead we notice here a particular relation obtained by combining (\ref{g1gT}) and the e.o.m.~relation (\ref{EoMgT}).  
Taking the sum of  (\ref{g1gT}) and ${d \over dx}g_{1T}^{(1)}(x)$ from (\ref{EoMgT}), one obtains
\begin{eqnarray}
 g_T^q(x) = g_1^q(x) + \frac{d}{dx} g_{1T}^{(1),q}(x) - 2 \, {\cal P} 
  \int_{-1}^1 dx' \frac{G_{FT}^q(x,x')}{(x-x')^2}.  
  \label{LIRgT}
\end{eqnarray}
This is the relation known as a LIR in the literature~\cite{Belitsky:1997ay,Accardi:2009au} (see also Refs.~\cite{Tangerman:1994bb,Mulders:1995dh,Kundu:2001pk,Goeke:2003az}).  
We notice that the quark mass term vanishes in this LIR.  

The $\hat{J}$-operator relation among $h_1$, $h_L$, 
and $H_{FL}$ can be obtained by choosing $\Gamma=i\gamma_5\sigma_{\rho\alpha}$ 
in (\ref{nonlocal_dist}) and $\Gamma'=\gamma_5$ in (\ref{total_dist0})\,\cite{Kodaira:1998jn,Belitsky:1997ay}. 
Taking the matrix element of the operator identity in the lightcone limit
$y^2\to 0$, one obtains\footnote{One needs to keep the 
nucleon mass $P^2=M^2$ to obtain (\ref{h1hL}). See \cite{Kodaira:1998jn,Belitsky:1997ay} for details.}
\begin{eqnarray}
&&-x^2{d\over dx}\left( {h_L^q(x)\over x}\right)=2h_1^q(x)
-{m_q\over M}{dg_1^q(x)\over dx}\nonumber\\
&&\qquad\qquad +2{\cal P}
\int_{-1}^1dx'{1\over x-x'}
\left({\partial \over \partial x}- {\partial \over \partial x'}
\right)H_{FL}^q(x,x')\,.
\label{h1hL}
\end{eqnarray}
Similar to the procedure that led from (\ref{g1gT}) to (\ref{LIRgT}), 
combining (\ref{h1hL}) and the e.o.m.~relation (\ref{EoMhL}) leads to another LIR, 
\begin{eqnarray}
 h_L^q(x) = h_1^q(x) - \frac{d}{dx} h_{1L}^{\perp (1),q}(x)
  + 2 \, {\cal P} \int_{-1}^1 dx' \frac{H_{FL}^q(x,x')}{(x-x')^2}.
 \label{LIRhL}
\end{eqnarray}
As in the case of (\ref{LIRgT}), this relation is independent of the quark mass. 

The choice of $\Gamma=1$ in (\ref{nonlocal_dist}) leads to a $\hat{J}$-operator relation between
$e$ and $H_{FU}$, but this is simply the same as the e.o.m.~relation
(\ref{EoMe}).  

\subsubsection{LIRs for twist-3 fragmentation functions}

In order to derive LIRs for the twist-3 fragmentation functions,
we need to extend the technique used in the previous subsection
to the non-local operators for the fragmentation matrix elements\,\cite{Balitsky:1990ck}.  
To this end we first define the following quantity, 
\begin{eqnarray}
 \langle\hat{I}^{[\Gamma]}_{\rho\alpha}(y)\rangle \!\!&\equiv &\!\!y_\rho \frac{\partial}{\partial y^\alpha}
\bra{0} \, {\cal W}[\pm\infty y;-y] \, q(-y)
 \, \ket{P_h,S_h;X} \nn\\
 && \times \bra{P_h,S_h;X} \, \bar{q}(y) \, \Gamma \, 
  {\cal W}[y;\pm\infty y] \, \ket{0}\,,
\label{Ihat}
\end{eqnarray}
where $\Gamma$ is an appropriate Dirac matrix.  
We need to relate  $ \langle\hat{I}^{[\Gamma]}_{\rho\alpha}(y)\rangle $ to the correlation functions
containing the field strength. 
The derivative in (\ref{Ihat}) can be performed using the identities
\begin{eqnarray}
&&\hspace{-0.3cm}{\partial\over \partial y^\mu}{\cal W}[\pm\infty y; -y] q(-y)
=-{\cal W}[\pm \infty y;y]D_\mu(-y) q(-y)\nonumber\\
&&\hspace{-0.3cm}\qquad+\.i\int_{-1}^{\pm\infty}\,dt\,t\,{\cal W}[\pm\infty y;ty]gF_{\mu y}(ty){\cal W}[ty;-y] q(-y)\,, 
\label{derivq}
\end{eqnarray}
and
\begin{eqnarray}
&&{\partial \over \partial y^\mu}\bar{q}(y){\cal W}[y;\pm\infty y]
=\bar{q}(y)\overleftarrow{D}_\mu(y){\cal W}[y;\pm\infty y]\nonumber\\
&&\qquad
+i\int_{\pm\infty}^1\,dt\,t\,\bar{q}(y) {\cal W}[y;t y] gF_{\mu y}(ty){\cal W}[ty;\pm\infty y]\,, 
\label{derivqbar}
\end{eqnarray}
where we have discarded  the boundary term at $\pm \infty y$. 
We also note that the total derivative
of the fragmentation correlator vanishes due to translational invariance,
\begin{eqnarray}
&&0= 
\partial_\rho
\bra{0} \, {\cal W}[\pm\infty y;-y] \, q(-y)
 \, \ket{P_h,S_h;X} \nn\\
 &&\qquad \times \bra{P_h,S_h;X} \, \bar{q}(y) \, \Gamma' \, 
  {\cal W}[y;\pm\infty y] \, \ket{0} \nonumber\\[5pt]
&&=
\lim_{\xi^\rho\to 0} {d\over d\xi^\rho}
\bra{0} \, {\cal W}[\pm\infty y+\xi;-y+\xi] \, q(-y+\xi)
 \, \ket{P_h,S_h;X} \nn\\
 &&\qquad \times \bra{P_h,S_h;X} \, \bar{q}(y+\xi) \, \Gamma' \, 
  {\cal W}[y+\xi;\pm\infty y+\xi] \, \ket{0} \nonumber\\[5pt]
&&
=
\bra{0} \, {\cal W}[\pm\infty y;-y] \, D_\rho (-y) q(-y)
 \, \ket{P_h,S_h;X} \nn\\
 &&\qquad \times \bra{P_h,S_h;X} \, \bar{q}(y) \, \Gamma' \, 
  {\cal W}[y;\pm\infty y] \, \ket{0} \nonumber\\[5pt]
&&+
\bra{0} \, {\cal W}[\pm\infty y;-y] \,  q(-y)
 \, \ket{P_h,S_h;X} \nn\\
 &&\qquad \times \bra{P_h,S_h;X} \, \bar{q}(y)\,\overleftarrow{D}_\rho (y) \, \Gamma' \, 
  {\cal W}[y;\pm\infty y] \, \ket{0} \nonumber\\[5pt]
&&+
\bra{0} \int_{-1}^{\pm\infty} dt\, {\cal W}[\pm\infty y;ty] ig F_{\rho y}(ty)  {\cal W}[ty;-y] q(-y)\nonumber\\
&&\qquad \times\ket{P_h,S_h;X} 
 \bra{P_h,S_h;X} \, \bar{q}(y) \, \Gamma' \, 
  {\cal W}[y;\pm\infty y] \, \ket{0} \nonumber\\[5pt]
&&+
\bra{0} {\cal W}[\pm\infty y;-y] q(-y)
 \, \ket{P_h,S_h;X} \nn\\
 && \times \bra{P_h,S_h;X}  \int^{1}_{\pm\infty} dt\, \bar{q}(y) \, \Gamma' \, 
 {\cal W}[y;ty]  ig F_{\rho y}(ty) 
  {\cal W}[ty;\pm\infty y] \, \ket{0}\,. \nonumber\\
  \label{total_frag}
\end{eqnarray}
Since the form of the gauge-links appearing in the fragmentation
matrix elements in Eqs.~(\ref{Ihat})--(\ref{total_frag}) is obvious, we shall omit them in the matrix elements below.  
As for the case of the distribution functions, the relation (\ref{total_frag}) can be used to
eliminate the terms which appear with the covariant derivative in $\langle\hat{I}^{[\Gamma]}_{\rho\alpha}(y)\rangle $
in favor of those with the field strength tensor.  

Unlike the distribution functions, dynamical twist-3 fragmentation functions 
($\hat{D}_{FT}$, $\hat{G}_{FT}$, $\hat{H}_{FU}$, and $\hat{H}_{FL}$) 
are complex, 
and the real parts of these functions are ``naively"
time-reversal even ($T$-even). Since their Lorentz and spinor structures are the same as those for the $T$-even distribution functions, 
the LIRs involving $\Re [\hat{G}_{FT}]$, $\Re [\hat{H}_{FL}]$
can be obtained  in parallel with the corresponding twist-3 distribution 
functions.  Using the same $\Gamma$ and $\Gamma'$ in (\ref{Ihat}) and (\ref{total_frag}), the LIRs read   
\begin{eqnarray}
 \frac{G_T^q(z)}{z} &=& \frac{G_1^q(z)}{z} 
  + \left( 1 - z\frac{d}{dz} \right) G_{1T}^{(1),q} (z) \nn\\
 && - \frac{2}{z}
  \int_z^\infty \frac{dz'}{z'^2}
  \frac{\Re [\hat{G}_{FT}^q(z,z')]}{(1/z-1/z')^2}, \label{LIRGT}\\[0.5cm]
 \frac{H_L^q(z)}{z} &=& \frac{H_1^q(z)}{z} - \left( 1 - z \frac{d}{dz} \right)
 H_{1L}^{\perp(1),q}(z) \nn\\
 && + \frac{2}{z} \int_z^\infty
 \frac{dz'}{z'^2}
 \frac{\Re [\hat{H}_{FL}^q(z,z')]}{(1/z-1/z')^2}, 
\label{LIRHL}
\end{eqnarray}
where we used the property $\frac{\partial}{\partial
(1/z')} \Delta_F^\rho(z,z') \vb_{z'=z}=0$, which is proven in Appendix \ref{App:AppendixA}. 
We note that (\ref{LIRGT}) 
can be obtained 
from (\ref{LIRgT}) 
by the formal replacement: $g_T(x) \to G_T(z)/z$,
$g_1(x) \to G_1(z)/z$, $G_{FT}(x,x')\to \Re [\hat{G}_{FT}(z,z')]/z$, 
$g_{1T}^{(1)}(x)\to G_{1T}^{(1)}(z)/z$ and ${d\over dx}\to {d\over d(1/z)}$.  
Likewise, Eq.~(\ref{LIRHL}) can be obtained from (\ref{LIRhL})
by 
$h_L(x)\to H_L(z)/z$, $h_1(x)\to H_1(z)/z$, $H_{FL}(x,x')\to \Re [\hat{H}_{FL}(z,z')]/z$,
$h_{1L}^{\perp (1)}(x)\to H_{1L}^{\perp (1)}(z)/z$ and ${d\over dx}\to {d\over d(1/z)}$.  

The derivation of LIRs involving the imaginary part of the dynamical twist-3
fragmentation functions requires new calculations.
For the relations among $D_T$, $\Im[D_{FT}]$, and $\Im[ G_{FT}]$,
we choose $\Gamma=\epsilon^{\alpha\mu\rho S_\perp} \gamma_\mu$
in (\ref{Ihat}) and 
$\Gamma'=
\epsilon_{\alpha \mu y S_\perp}(\sigma^{\alpha\mu}\gamma^\rho + \gamma^\rho\sigma^{\alpha\mu})$ in (\ref{total_frag}). 
One then obtains the following relation:
\begin{eqnarray}
&&\epsilon^{\alpha\mu y S_\perp} {\partial \over \partial y^\alpha}
\bra{0} 
\gamma_\mu q(-y)\ket{P_h,S_h;X} 
\bra{P_h,S_h;X} \bar{q}(y)
\ket{0}\nonumber\\[5pt]
&& =\bra{0}\int_{-1}^{\pm\infty}  
dt\, 
g F_{S_\perp y}(ty)\gamma_5 \slash y 
q (-y) \ket{P_h,S_h;X}   \bra{P_h,S_h;X}\bar{q}(y) 
\ket{0}
\nonumber\\[5pt]
&&+\bra{0} 
\gamma_5 \slash y q (-y) \ket{P_h,S_h;X}
 \bra{P_h,S_h;X} \int_{\pm\infty}^1 dt\, \bar{q}(y) 
g F_{S_\perp y}(ty) 
\ket{0}\nonumber\\[5pt]
&&+\epsilon^{\alpha \mu y  S_\perp}
\bra{0}\int_{-1}^{\pm\infty} dt\,t\, 
 igF_{\alpha y}(ty)  
\gamma_\mu q(-y) \ket{P_h,S_h;X} \nonumber\\
&&\qquad\qquad\times \bra{P_h,S_h;X} \bar{q}(y)
\ket{0}\nonumber\\[5pt]
&&+ \epsilon^{\alpha \mu y  S_\perp}
\bra{0} 
\gamma_\mu q(-y) \ket{P_h,S_h;X}\nonumber\\
&&\qquad\qquad\times \bra{P_h,S_h;X} \int_{\pm\infty}^1 dt\,t\, \bar{q}(y) 
ig F_{\alpha y}(ty)   
\ket{0}\,, \label{e:Ihatlong}
\end{eqnarray}
where we found that the quark mass terms cancel out.  
Taking the lightcone limit ($y^2\to 0$) of (\ref{e:Ihatlong}) as well as the sum over $X$,
one obtains the following ``$\hat{I}$-operator relation'':
\begin{eqnarray}
&&{-1\over z}{d\over d(1/z)}\left( {D_T^q(z)\over z}\right)\nonumber\\
&&\qquad=\int d\left({1\over z_1}\right)
{\cal P}\left( {1\over 1/z-1/z_1}\right)\nonumber\\
&&\qquad\qquad \times 
\left\{
\left( {\partial \over \partial (1/z)} + {\partial\over \partial (1/z_1)}\right) 
{\Im [\hat{G}_{FT}^q(z,z_1)]\over z}\right.\nonumber\\
&&\left.
\qquad\qquad
-\left( {\partial \over \partial (1/z)} - {\partial\over \partial (1/z_1)}\right) 
{\Im [\hat{D}_{FT}^q(z,z_1)]\over z}
\right\}.  
\label{DTWW}
\end{eqnarray}
This is a new relation derived here for the first time. 
By taking the sum of (\ref{DTWW}) and ${d\over
d(1/z)}\left({1\over z}D_{1T}^{\perp (1),q}(z)\right)$ from
(\ref{EoMDT}), one obtains the LIR as
\begin{eqnarray}
\frac{D_T^q(z)}{z} &=& - \left( 1- z \frac{d}{dz} \right) D_{1T}^{\perp
(1),q}(z) \nn\\
 && - \frac{2}{z} \int_z^\infty \frac{dz'}{z'^2}
 \frac{\Im [\hat{D}_{FT}^q(z,z')]}{(1/z-1/z')^2}. \label{LIRDT}
\end{eqnarray}

For the $\hat{I}$-operator relation among $H$ and $\Im[\hat{H}_{FU}]$, we choose
$\Gamma=\epsilon^{\rho\alpha\mu\nu}\sigma_{\mu\nu}i\gamma_5$ in (\ref{Ihat}) 
and $\Gamma'=1$ in (\ref{total_frag}), which leads to the following identity:
\begin{eqnarray}
&&\epsilon^{\mu\nu\rho\alpha}y_\rho {\partial\over \partial y^\alpha}
\bra{0} q(-y)  \ket{P_h,S_h;X} \bra{P_h,S_h;X}
\bar{q}(y)\sigma_{\mu\nu}i\gamma_5 \ket{0}\nonumber\\
&&=\epsilon^{\mu\nu y \alpha}
\bra{0}\int_{-1}^{\pm\infty} dt\,t\,ig\,F_{\alpha y}(ty) q(-y) \ket{P_h,S_h;X}\nonumber\\
&&\qquad\qquad\times
\bra{P_h,S_h;X} \bar{q}(y)\sigma_{\mu\nu}\,i\gamma_5 \ket{0}\nonumber\\
&&+\epsilon^{\mu\nu y \alpha}
\bra{0} q(-y) \ket{P_h,S_h;X}\nonumber\\
&&\qquad\times\bra{P_h,S_h;X} \int_{\pm\infty}^1 dt\,t\,\bar{q}(y)
igF_{\alpha y}(ty)\,i\sigma_{\mu\nu}\gamma_5 \ket{0},
\end{eqnarray}
where the quark-mass terms again cancel out. 
Taking the lightcone limit of this equation, one arrives at
\begin{eqnarray}
{1\over z^2} {d H^q(z)\over d(1/z)} &=& 2 {\cal P}\int d\left({1\over z_1}\right)
{1\over 1/z_1-1/z } \nonumber\\
&&\times\left( {\partial \over \partial\left(1/z_1\right) } - {\partial \over \partial 
\left( 1/z\right)}
\right) {\Im[\hat{H}_{FU}^q(z,z_1)]\over z}.  
\label{HHFU}
\end{eqnarray}
This result is also new.  
As in (\ref{LIRDT}), 
combining (\ref{HHFU}) with the e.o.m.~relation (\ref{EoMH})
leads to
\begin{eqnarray}
 \frac{H^q(z)}{z} &=& - \left( 1 - z \frac{d}{dz} \right) H_{1}^{\perp
  (1),q}(z) \nn\\
 && - \frac{2}{z} \int_z^\infty \frac{dz'}{z'^2}
  \frac{\Im [\hat{H}_{FU}^q(z,z')]}{(1/z-1/z')^2}. \label{LIRH}
\end{eqnarray}
This completes the derivation of all LIRs for twist-3 distribution functions and fragmentation functions. 

\subsection{Intrinsic and kinematical twist-3 functions expressed in terms of dynamical twist-3 functions}

In this subsection we show that the use of the e.o.m.~relations and the $\hat{J}$- and $\hat{I}$-operator relations from the previous subsections allows one to express the intrinsic and kinematical twist-3 functions in terms of the
dynamical twist-3 functions and the usual twist-2 parton distribution/fragmentation functions.
Omitting the former in these expressions provides a generalized Wandzura-Wilczek (WW) approximation~\cite{Wandzura:1977qf}. 
Since the dynamical twist-3 functions and the twist-2 functions do not mix under renormalization,
these formulae provide a basis to derive the scale dependence of the twist-3 functions.  
The evolution equations for the dynamical and the intrinsic twist-3 functions 
have been derived in the literature (see~\cite{Balitsky:1987bk,Ali:1991em,Koike:1994st,Kodaira:1996md,Balitsky:1996uh,Koike:1996bs,Belitsky:1997zw,Belitsky:1997by,Kang:2008ey,Vogelsang:2009pj,Zhou:2008mz,Braun:2009mi,Kang:2010xv,Ma:2011nd,Zhou:2015lxa} and references therein).  
It is also clear from the relations below that the dynamical twist-3 functions constitute a complete set
of twist-3 correlators and all twist-3 cross sections in collinear factorization
can be written in terms of twist-2 functions and the dynamical twist-3 functions.  
In this sense the intrinsic and kinematical twist-3 correlators can be viewed as auxiliary functions.  The use of those functions, however,
sometimes leads to a concise expression for twist-3 cross sections, at least in LO calculations,
which makes them convenient to keep for phenomenological analyses.

Here we derive the explicit expressions for the intrinsic and kinematical twist-3 functions
in terms of the twist-2 functions and the dynamical twist-3 functions
using the e.o.m.~relations and the $\hat{J}$- and $\hat{I}$-operator relations obtained in the previous subsections.
First, the integration of (\ref{g1gT}) immediately gives
\begin{eqnarray}
g_T^q(x)&=&\int_x^{\epsilon(x)}dx'\,\frac{g_1^q(x')}{x'}
+{m_q\over M}\left( {1\over x}h_1^q(x) +\int^x_{\epsilon(x)}dx' \,{h_1^q(x')\over {x'}^2}\right)\nonumber\\
&&\hspace{0cm}+ \int_x^{\epsilon(x)}\frac{dx_1}{x_1^2}\,\mathcal{P}\int_{-1}^1dx_2\,\Bigg[\frac{1-x_1\,\delta(x_1-x)}{x_1-x_2}F_{FT}^q(x_1,x_2)\nonumber\\
&&-\frac{3x_1-x_2-x_1(x_1-x_2)\,\delta(x_1-x)\,}{(x_1-x_2)^2}G_{FT}^q(x_1,x_2)\Bigg]\, ,\nonumber\\
\label{gT}
\end{eqnarray}
where $\epsilon(x)=2\theta(x)-1$, and $g_T^q(x)$ at $x<0$ represents the antiquark distribution as stated in Sec.~\ref{s:par}.  
This is the WW relation, which decomposes the twist-3 quark distribution $g_T$ into the contributions from
$g_1$ and the dynamical twist-3 distributions $F_{FT}$ and $G_{FT}$ (and a quark mass term).  Note that we have used the delta function $\delta(x_1-x)$ to make the expression (\ref{gT}) more compact, and it is to be understood that $x$ falls within the range of integration $(x,\epsilon(x))$, i.e.,  $\int_x^{\epsilon(x)} d{x_1}f(x_1)\delta(x_1-x) = f(x)$.
Inserting Eq.~(\ref{gT}) into the e.o.m.~relation (\ref{EoMgT}), one obtains
\begin{eqnarray}
g_{1T}^{(1),q}(x)& = &x\int_x^{\epsilon(x)}dx'\,\frac{g_1^q(x')}{x'} +{m_q\over M} x \int^x_{\epsilon(x)}dx'\,{h_1^q(x')\over {x'}^2}\label{g1T}\\
&&\hspace{0cm}+ x\int_x^{\epsilon(x)}\frac{dx_1}{x_1^2}\,\mathcal{P}\int_{-1}^1dx_2\,\Bigg[\frac{F_{FT}^q(x_1,x_2)}{x_1-x_2}\nonumber\\
&&-\frac{(3x_1-x_2)\,G_{FT}^q(x_1,x_2)}{(x_1-x_2)^2}\Bigg]\,.\nonumber
\end{eqnarray}
This result shows that the $g_{1T}^{(1)}$ can also be written in terms of
$g_1$ and the dynamical twist-3 distributions $F_{FT}$ and $G_{FT}$.  

Likewise, integration of (\ref{h1hL}) gives
\begin{eqnarray}
&&\!\!\!\!h_L^q(x)=2x\int_x^{\epsilon(x)}dx_1\,{h_1^q(x_1)\over x_1^2}\nonumber\\
&&\!\!\!\!\!+{m_q\over M}\left( {g_1^q(x)\over x} -2x \int_x^{\epsilon(x)}dx_1\,{g_1^q(x_1)\over x_1^3}\right)\nonumber\\
&&\!\!\!\!\!+4x \int_x^{\epsilon(x)}{dx_1\over x_1^3}\times\nonumber\\ 
&&{\cal P}\int_{-1}^1dx_2\,{(x_1/2)(x_2-x_1)\delta(x_1-x)+2x_1-x_2 \over (x_1-x_2)^2}H_{FL}^q(x_1,x_2)\,.\nonumber\\ 
\label{hLWW}
\end{eqnarray}
This is the WW analogue for $h_L$, which shows that
$h_L$ can be expressed in terms of the twist-2 distribution $h_1$ and the dynamical twist-3 distribution $H_{FL}$.  
Analogous to the case of $g_T$ and $g_{1T}^{\perp (1)}$,
insertion of (\ref{hLWW}) into (\ref{EoMhL}) leads to the expression for $h_{1L}^{\perp (1)}$ as
\begin{eqnarray}
&&h_{1L}^{\perp (1),q}(x)=x^2\int_x^{\epsilon(x)}dx_1\,{h_1^q(x_1)\over x_1^2}\nonumber\\
&&-{m_q\over M}x^2 \int_x^{\epsilon(x)}dx_1\,{g_1^q(x_1)\over x_1^3}\nonumber\\
&&+2x^2 \int_x^{\epsilon(x)}{dx_1\over x_1^3}\,{\cal P}\int_{-1}^1dx_2\,{2x_1-x_2 \over (x_1-x_2)^2}H_{FL}^q(x_1,x_2)\,.  
\label{h1L(1)WW}
\end{eqnarray}

One can repeat the same procedure for the fragmentation functions.
From (\ref{HHFU}) and (\ref{EoMH}), one obtains the results for $H$ and
$H_1^{\perp (1)}$ as
\begin{eqnarray}
H^{q}(z)& = & \int_z^1dz_1\int_{z_1}^\infty \frac{dz_2}{z_2^2}\,\times\nonumber\\
&& \hspace{-1.2cm}2\,\Bigg[\frac{\left(\,2(\frac{2}{z_1}-\frac{1}{z_2})+\frac{1}{z_1}\left(\frac{1}{z_1}-\frac{1}{z_2}\right)\,
\delta\left(\frac{1}{z_1}-\frac{1}{z}\right)\right)\,\Im[\hat{H}_{FU}^{q}(z_1,z_2)]}{\left(\frac{1}{z_1}-\frac{1}{z_2}\right)^2}\Bigg]\,,\nonumber\\\label{H}
\end{eqnarray}
\begin{eqnarray}
H_1^{\perp (1),q}(z)&=&-\tfrac{2}{z}\int_z^1dz_1\int_{z_1}^\infty \frac{dz_2}{z_2^2}\,\frac{\left(\frac{2}{z_1}-\frac{1}{z_2}\right)\,
\Im[\hat{H}_{FU}^{q}(z_1,z_2)]}{\left(\frac{1}{z_1}-\frac{1}{z_2}\right)^2}\,,\nonumber\\
&&\label{H1perp1}
\end{eqnarray}
where again with regards to the delta function $\delta(1/z_1-1/z)$ it is to be understood that $z$ falls within the range of integration $(z,1)$, i.e.,  $\int_z^{1} d{z_1}D(z_1)\delta(1/z_1-1/z) = z^2D(z)$.

The formulae for $D_T$ and $D_{1T}^{\perp (1)}$ can be obtained from (\ref{DTWW}) and (\ref{EoMDT}), and they read
\begin{eqnarray}
D_{T}^{q}(z)&=&-z\int_z^1\frac{dz_1}{z_1}\int_{z_1}^\infty \frac{dz_2}{z_2^2}\,\times\nonumber\\
&&\hspace{-0.5cm}\Bigg[\frac{\left(1+\frac{1}{z_1}\delta\left(\frac{1}{z_1}-\frac{1}{z}\right)\right)\,\Im[\hat{G}_{FT}^{q}(z_1,z_2)]}{\frac{1}{z_1}-\frac{1}{z_2}}\nonumber\\
&&\hspace{-0.5cm}-\frac{\left(\frac{3}{z_1}-\frac{1}{z_2}+\frac{1}{z_1}\left(\frac{1}{z_1}-\frac{1}{z_2}\right)\,\delta\left(\frac{1}{z_1}
-\frac{1}{z}\right)\right)\,\Im[\hat{D}_{FT}^{q}(z_1,z_2)]}{\left(\frac{1}{z_1}-\frac{1}{z_2}\right)^2}\Bigg]\,,\nonumber\\ \label{DT}
\end{eqnarray}
\begin{eqnarray}
D_{1T}^{\perp (1),q}(z)&=&\int_z^1\frac{dz_1}{z_1}\int_{z_1}^\infty \frac{dz_2}{z_2^2}\,
\Bigg[\frac{\Im[\hat{G}_{FT}^{q}(z_1,z_2)]}{\frac{1}{z_1}-\frac{1}{z_2}}\nonumber \\
&&-\,\frac{\left(\frac{3}{z_1}-\frac{1}{z_2}\right)\,\Im[\hat{D}_{FT}^{q}(z_1,z_2)]}{\left(\frac{1}{z_1}-\frac{1}{z_2}\right)^2}\Bigg]\,.\label{D1Tperp1}
\end{eqnarray}

The formulae for $G_T$ and $G_{1T}^{(1)}$ are given by
\begin{eqnarray}
G_{T}^{q}(z)&=&-z\int_z^1dz'\,\frac{G_1^q(z')}{z'^2}-z\int_z^1\frac{dz_1}{z_1}\int_{z_1}^\infty \frac{dz_2}{z_2^2}\,\times\nonumber\\
&&\hspace{-0.5cm}\Bigg[\frac{\left(1+\frac{1}{z_1}\delta\left(\frac{1}{z_1}-\frac{1}{z}\right)\right)\,
\Re[\hat{D}_{FT}^{q}(z_1,z_2)]}{\frac{1}{z_1}-\frac{1}{z_2}}\nonumber\\
&&\hspace{-0.5cm}-\frac{\left(\frac{3}{z_1}-\frac{1}{z_2}+\frac{1}{z_1}\left(\frac{1}{z_1}
-\frac{1}{z_2}\right)\,\delta\left(\frac{1}{z_1}-\frac{1}{z}\right)\right)\,
\Re[\hat{G}_{FT}^{q}(z_1,z_2)]}{\left(\frac{1}{z_1}-\frac{1}{z_2}\right)^2}\Bigg]\nonumber\\
&&+{m_q\over M_h}\left(z\,H^q_1(z) + z\int_z^1dz'\,\frac{H^q_1(z')} {z'}\right)\, ,
\label{GT}
\end{eqnarray}
\begin{eqnarray}
G_{1T}^{(1),q}(z)&=&-\int_z^1dz'\,\frac{G_1^q(z')}{z'^2}-\int_z^1\frac{dz_1}{z_1}\int_{z_1}^\infty \frac{dz_2}{z_2^2}\times \nonumber\\
&&\hspace{-0.5cm}\Bigg[\frac{\Re[\hat{D}_{FT}^{q}(z_1,z_2)]}{\frac{1}{z_1}-\frac{1}{z_2}}-\frac{\left(\frac{3}{z_1}-\frac{1}{z_2}\right)\,\Re[\hat{G}_{FT}^{q}(z_1,z_2)]}{\left(\frac{1}{z_1}-\frac{1}{z_2}\right)^2}\Bigg]\nonumber\\
&&+{m_q\over M_h}\int_z^1dz'\,\frac{H_1^q(z')} {z'}\, ,
\label{G1T1}
\end{eqnarray}

Finally, the formulae for $H_L$ and $H_{1L}^{\perp (1)}$ read
\begin{eqnarray}
&&\!\!\!\!\!H_L^q(z)=-2\int_z^1 dz'\,{H_1^q(z')\over z'} +{m_q\over M_h} \left(z\,G_1^q(z)+2\int_z^1dz'\,G_1^q(z')\right)\nonumber\\
&&\!\!\!\!\! -\,4\int_z^1dz_1\int_{z_1}^\infty {dz_2\over z_2^2}\,\Re [\hat{H}_{FL}^q(z_1,z_2)]\times \nonumber\\
&&\quad{(1/(2z))(1/z_1-1/z_2)\delta(1/z_1-1/z)+2/z_1- 1/z_2 \over (1/z_1-1/z_2)^2 } \,, \nonumber\\
\label{HLWW}
\end{eqnarray}
\begin{eqnarray}
&&H_{1L}^{\perp (1),q}(z)=-{1\over z}\int_z^1dz'\,{H_1^q(z')\over z'}+{m_q\over M_h}{1\over z} \int_z^1dz'\,G_1^q(z')\nonumber\\
&&\quad -{2\over z}\int_z^1dz_1\int_{z_1}^\infty \, {dz_2\over z_2^2} {2/z_1 - 1/z_2 \over (1/z_1-1/z_2)^2}
\Re [ \hat{H}_{FL}^q(z_1,z_2) ]\,.
\label{H1LWW}
\end{eqnarray}

Like for their counterparts $g_T,\ g_{1T}^{(1)}$ and $h_L, h_{1L}^{\perp (1)}$ on the distribution side, we also 
have WW-approximations for the fragmentation functions $G_T,\ G_{1T}^{(1)}$ in Eqs.~(\ref{GT}),  (\ref{G1T1}) 
and $H_L,\ H_{1L}^{\perp (1)}$ in Eqs.~(\ref{HLWW}) and (\ref{H1LWW}). By dropping the quark-gluon-quark 
fragmentation functions $\hat{D}_{FT}$, $\hat{G}_{FT}$, and $\hat{H}_{FL}$, one can describe the 
functions $G_{1T}^{(1)},\ G_T$ and $H_L,\ H_{1L}^{\perp (1)}$ solely by the twist-2 fragmentation 
functions $G_1$ and $H_1$, respectively.  We note that the other intrinsic and kinematical 
twist-3 functions found in Eqs.~(\ref{H})--(\ref{D1Tperp1}), like the moment $H_1^{\perp(1)}$ of the Collins 
function, vanish if one neglects three-parton fragmentation functions.

\section{Lorentz invariance in the calculation of partonic coefficients for \boldmath{$\ell N\to \lowercase{h} X$} in collinear twist-3 factorization\label{s:LI}}

In this section we demonstrate that the LIRs in Eqs.~(\ref{LIRgT}), (\ref{LIRH}), (\ref{LIRDT}), and (\ref{LIRGT}) not only serve as constraints on the various twist-3 functions but are of vital importance in the calculation of physical processes. To this end we focus on the ``simple" reaction, $\ell(l) + N(P)\to h(P_h)+X$, where the collinear twist-3 formalism can be easily applied. We re-evaluate several transverse spin observables in this process at LO accuracy and show explicitly how LIRs are used in order to obtain theoretically consistent results.

One observation about the Lorentz invariance in the twist-3 formalism has already been made in Ref.~\cite{Kanazawa:2014tda}.  In that work, the double-spin asymmetry $A_{LT}$ of a longitudinally polarized lepton colliding with a transversely polarized nucleon was calculated, i.e., $\vec{\ell}\,N^{\uparrow}\to h\,X$. In fact, this double-spin observable was computed in two frames: in the center-of-mass frame of the lepton and nucleon (see Eq.~(17) of Ref.~\cite{Kanazawa:2014tda}), and in the center-of-mass frame of the nucleon and hadron (see Eq.~(18) of Ref.~\cite{Kanazawa:2014tda}). 
While the twist-3 fragmentation term for $A_{LT}$ was manifestly the same in both frames, the contribution from twist-3 distribution functions in the two frames initially seemed to be different.  But with the help of the LIR (\ref{LIRgT}) this piece could be brought to the same form in both frames.

In general, twist-3 cross sections, {\it a priori}, could be frame-dependent since the partonic coefficient functions depend on the lightcone vectors $n$ and $m$, which in turn depend on the chosen reference frame.
The lightcone vectors enter through the parameterizations (\ref{eq:PhixParam}), (\ref{eq:ParamPhipart}), and (\ref{eq:ParamPhiF}) of the matrix elements on the nucleon side as well as in the parameterizations (\ref{eq:DeltazParam}), (\ref{eq:ParamDeltapart}), and (\ref{eq:ParamDeltaF}) on the fragmentation side.
It is only after applying the LIRs that one eliminates the frame-dependence and sees the Lorentz invariance of the result.
Note that an explicit dependence on the lightcone vectors does not show up in the calculations of partonic twist-2 cross sections but can be naturally expected for twist-3 observables because every description of a twist-3 observable will contain ``transverse" quantities (e.g., a transverse spin component).
And, as mentioned before, one needs two lightcone directions to define a ``transverse" plane.

\begin{figure*}[t]
\centering
\subfloat[]{\includegraphics[width=0.30\textwidth]{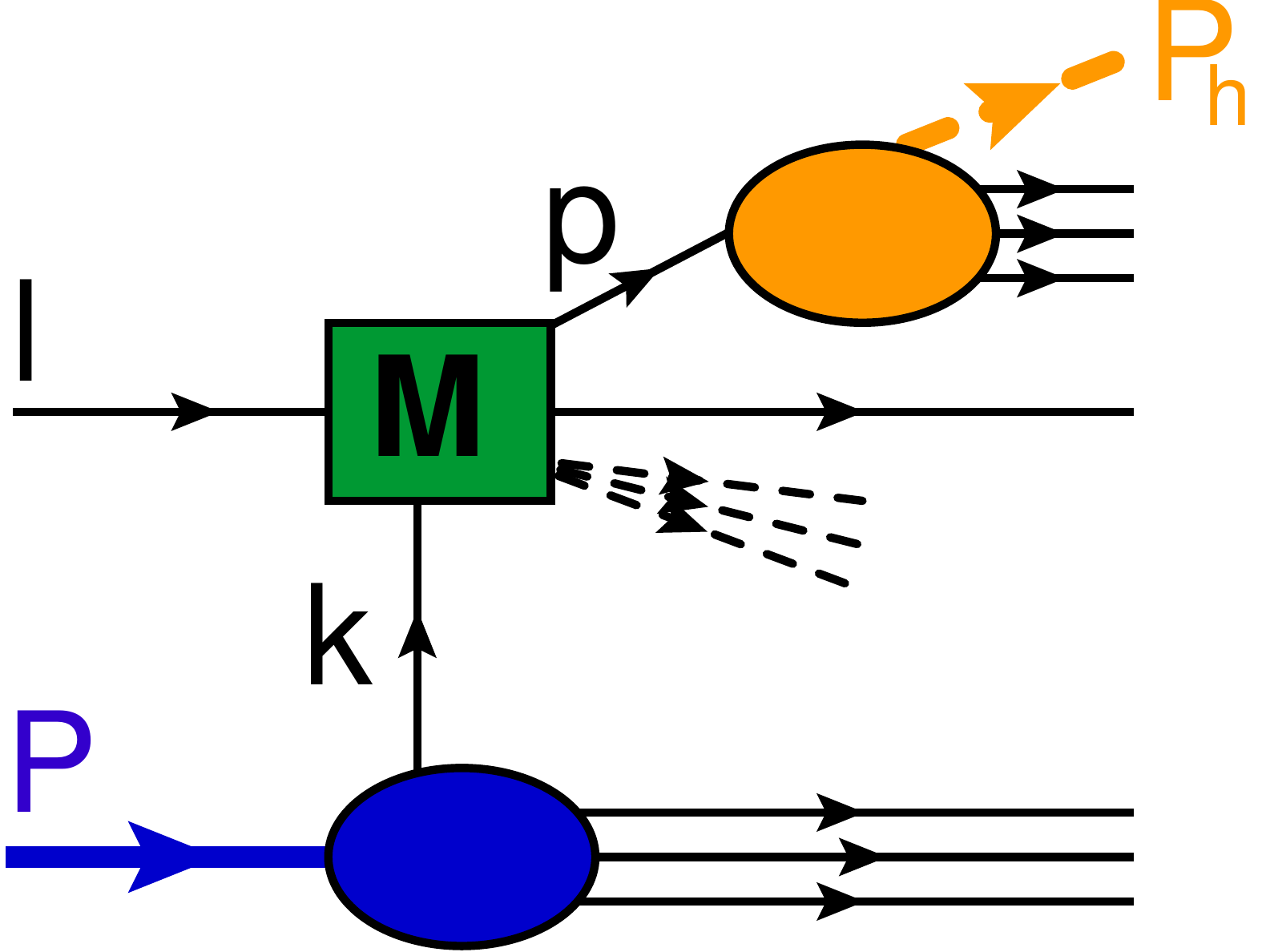}\label{fig:q2q}}\hspace{1.5cm}
\subfloat[]{\includegraphics[width=0.30\textwidth]{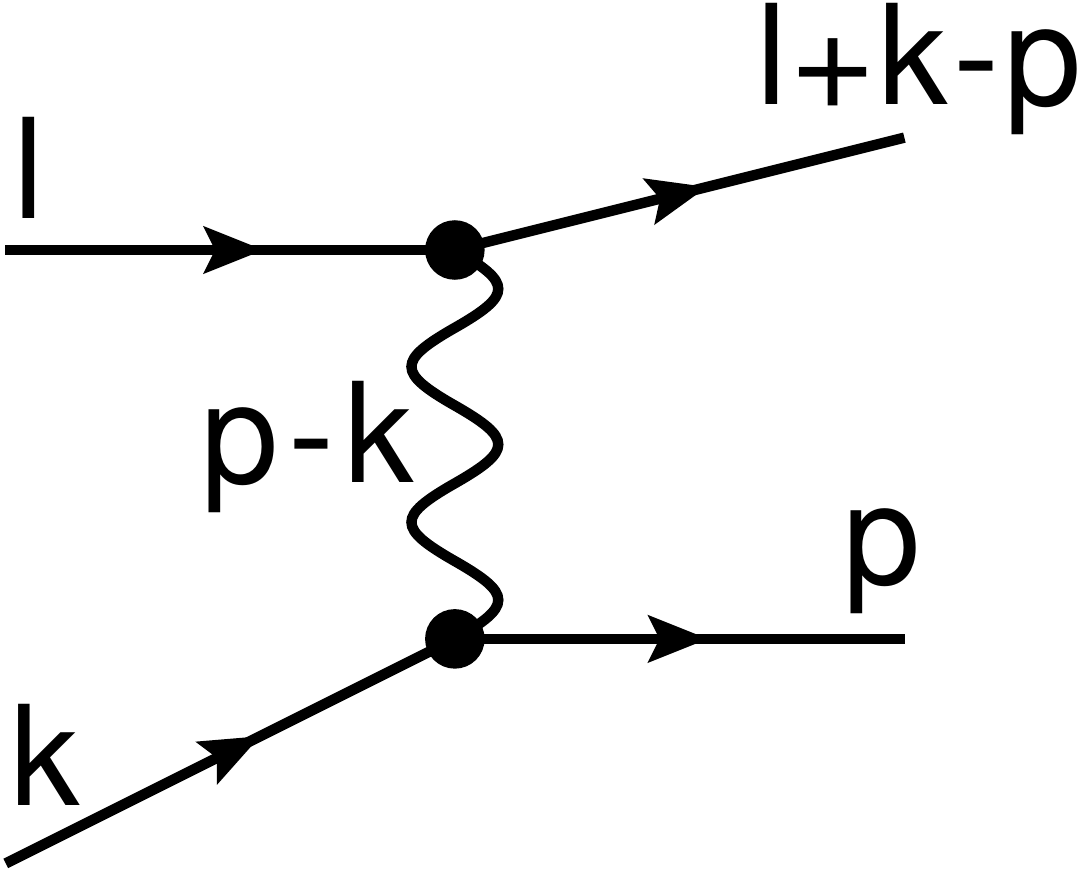}\label{fig:q2qLO}}

\caption{{Subfigure (a):} Factorization (of the amplitude) in terms of two-parton quark correlators of the process $\ell (l) + N(P)\to h(P_h)+X$. The square blob stands for the hard partonic scattering while the bottom and top oval blobs represent the soft distribution and fragmentation functions, respectively. This diagram describes twist-2 observables and intrinsic and kinematical twist-3 effects. {Subfigure (b):} Leading-order diagram for the hard partonic scattering. \label{fig:q2qFIG}}
\end{figure*}

The lightcone vectors $n$ and $m$ can always be expressed in terms of physical vectors that are inherent to the process under consideration.
For a general inclusive process $A(P_A)+B(P_B)\to C(P_C)+X$ one might utilize the momenta $P_A$, $P_B$, and $P_C$ of particles $A$, $B$, and $C$, respectively. In our example process, $\ell(l)+N(P)\to h(P_h)+X$, we express the lightcone momenta in terms of $l^\mu$, $P^\mu$, $P_h^\mu$, and a fourth 4-vector $\epsilon^{\mu\nu\rho\sigma}l_\nu P_\rho P_{h\sigma}\equiv\epsilon^{\mu lPP_h}$. Those vectors form a basis of the Minkowski space, and we can write the lightcone vectors $n$ and $m$ in the most general way as
\begin{eqnarray}
n^\mu & = & \chi_P\,P^\mu+\chi_l\,l^\mu+\chi_h\,P_h^\mu+\chi_\epsilon\, \epsilon^{\mu l P P_h},\nonumber\\
m^\mu& = & \eta_P\,P^\mu+\eta_l\,l^\mu+\eta_h\,P_h^\mu+\eta_\epsilon\,\epsilon^{\mu l P P_h}.\label{DecompLCVectors}
\end{eqnarray}
Such decompositions like Eq.~(\ref{DecompLCVectors}) are valid for any 4-vector. In addition, the vectors $n^\mu$ and $m^\mu$ must obey the constraints $n^2=m^2=0$ and $P\cdot n=P_h\cdot m=1$. These conditions eliminate four of the eight coefficients, and we obtain the following linear combinations,
\begin{eqnarray}
n^\mu& = & \chi\,P_h^\mu+\tfrac{2+\chi T}{S}\,l^\mu+\left[\tfrac{\chi U(2+\chi T)}{2S}+\tfrac{S T U}{8}\chi_\epsilon^2\right]\,P^\mu+\chi_\epsilon\,\epsilon^{\mu lPP_h},\nonumber\\
m^\mu&=& \eta\,P^\mu-\tfrac{2+\eta T}{U}\,l^\mu+\left[\tfrac{\eta S(2+\eta T)}{2U}+\tfrac{S T U}{8}\eta_\epsilon^2\right]\,P_h^\mu+\eta_\epsilon \epsilon^{\mu lPP_h}.\nonumber\\ \label{LCVectors}
\end{eqnarray}
In these expression we have made use of the Mandelstam variables $S=(l+P)^2$, $T=(P-P_h)^2$, and $U=(l-P_h)^2$. We note that the coefficients $\chi$, $\eta$, $\chi_\epsilon$, and $\eta_\epsilon$ are left unconstrained, and any choice of those coefficients may be acceptable unless one imposes additional requirements. 
In fact, different values for these coefficients correspond to different choices of reference frames.
In the case of $\ell \,N\to h\,X$, for instance, one may work in the lepton--nucleon c.m.~frame, which is often a reasonable approximation to the lab frame. 
In this frame, the lightcone vector $n^\mu$ would be a vector that is collinear to $l^\mu$. In (\ref{LCVectors}) this would correspond to $\chi = 0 $ and $\chi_\epsilon=0$.
On the other hand, if one chooses to work in a nucleon--hadron c.m.~frame (often a convenient choice for calculations), one selects lightcone vectors $n^\mu$ and $m^\mu$ which are collinear to $P_h^\mu$ and $P^\mu$, respectively. Hence, one has $\chi =\eta=-2/T$ and $\chi_\epsilon = \eta_\epsilon = 0$ for this choice of frame.
Consequently, one sees in the collinear twist-3 formalism that there is the potential to obtain seemingly different results in different frames (as in Ref.~\cite{Kanazawa:2014tda}) due to the fact that different lightcone vectors $n^\mu$ and $m^\mu$ are used.  However, as was the case in Ref.~\cite{Kanazawa:2014tda}, and as we will see below, the use of LIRs resolves the apparent discrepancy.

We will proceed with a re-evaluation of four transverse spin observables in the inclusive hadron production process in lepton-nucleon scattering $\ell \,N\to h\,X$, using now the general form of the lightcone vectors $n^\mu$ and $m^\mu$ in (\ref{LCVectors}).  
We emphasize that leaving the coefficients $\chi$, $\eta$, $\chi_\epsilon$, $\eta_\epsilon$ unconstrained corresponds to a calculation for an {\it arbitrary} reference frame.
The reactions we will consider are 1) The SSA of a transversely polarized nucleon $A_{UTU}$; 2) the DSA of a longitudinally polarized lepton and a transversely polarized nucleon $A_{LTU}$; 3) the SSA of a transversely polarized hadron $A_{UUT}$ (e.g., $\Lambda$-hyperon production); and 4) the DSA of a longitudinally polarized lepton and a transversely polarized hadron $A_{LUT}$. The SSA $A_{UTU}$ has been calculated in Ref.~\cite{Gamberg:2014eia} at LO accuracy. A LO calculation for the DSA $A_{LTU}$ has been presented in Ref.~\cite{Kanazawa:2014tda}, and recently the SSA for polarized $\Lambda$-hyperons was calculated in Ref.~\cite{Kanazawa:2015jxa}. To the best of our knowledge the double-spin asymmetry $A_{LUT}$ of a longitudinally polarized lepton and a transversely polarized outgoing hadron has never been analyzed in the collinear twist-3 formalism. 

\begin{figure*}[t]
\centering
\subfloat[]{\includegraphics[width=0.30\textwidth]{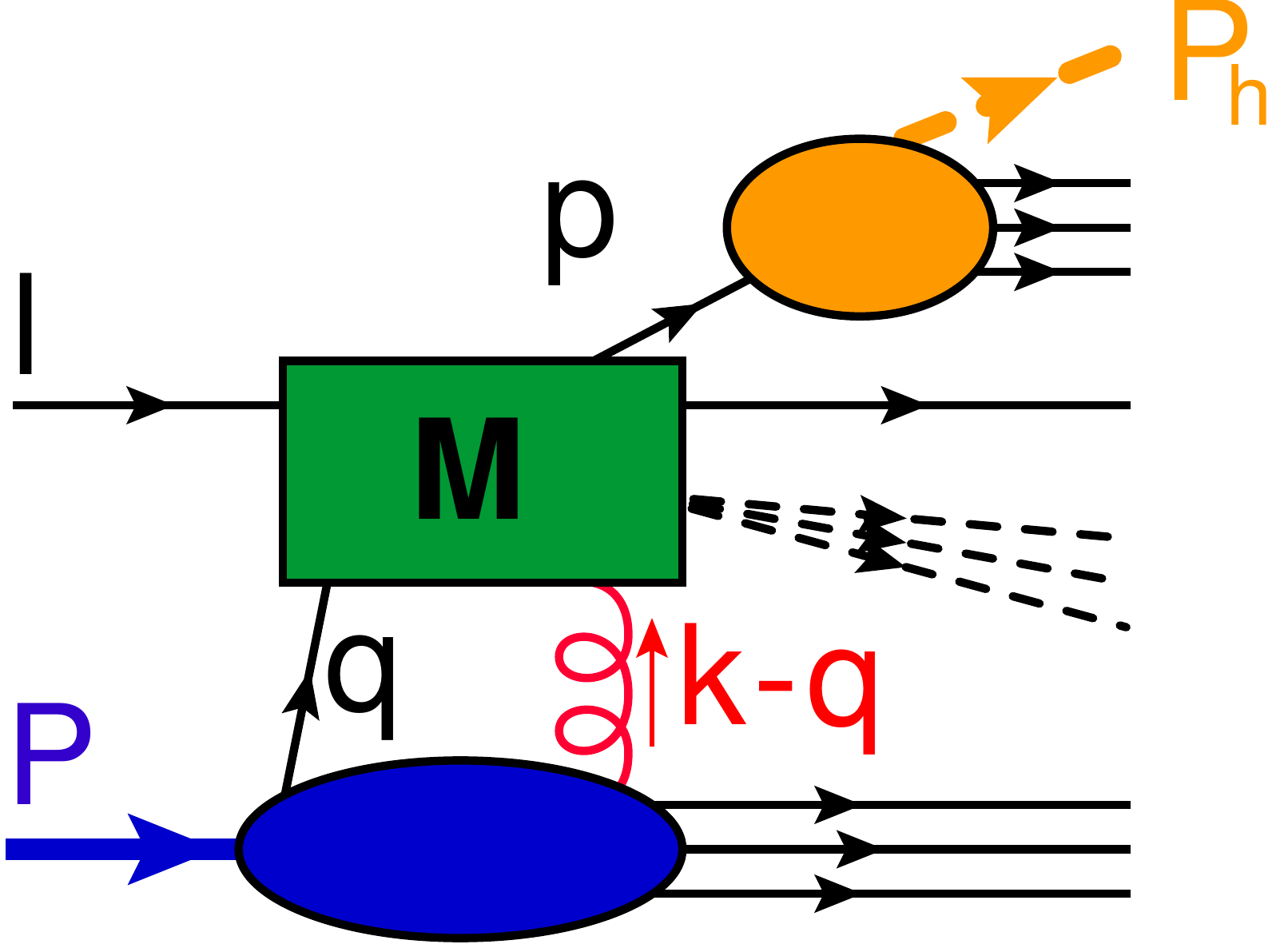}\label{fig:qg2q}}\hspace{1.5cm}
\subfloat[]{\includegraphics[width=0.30\textwidth]{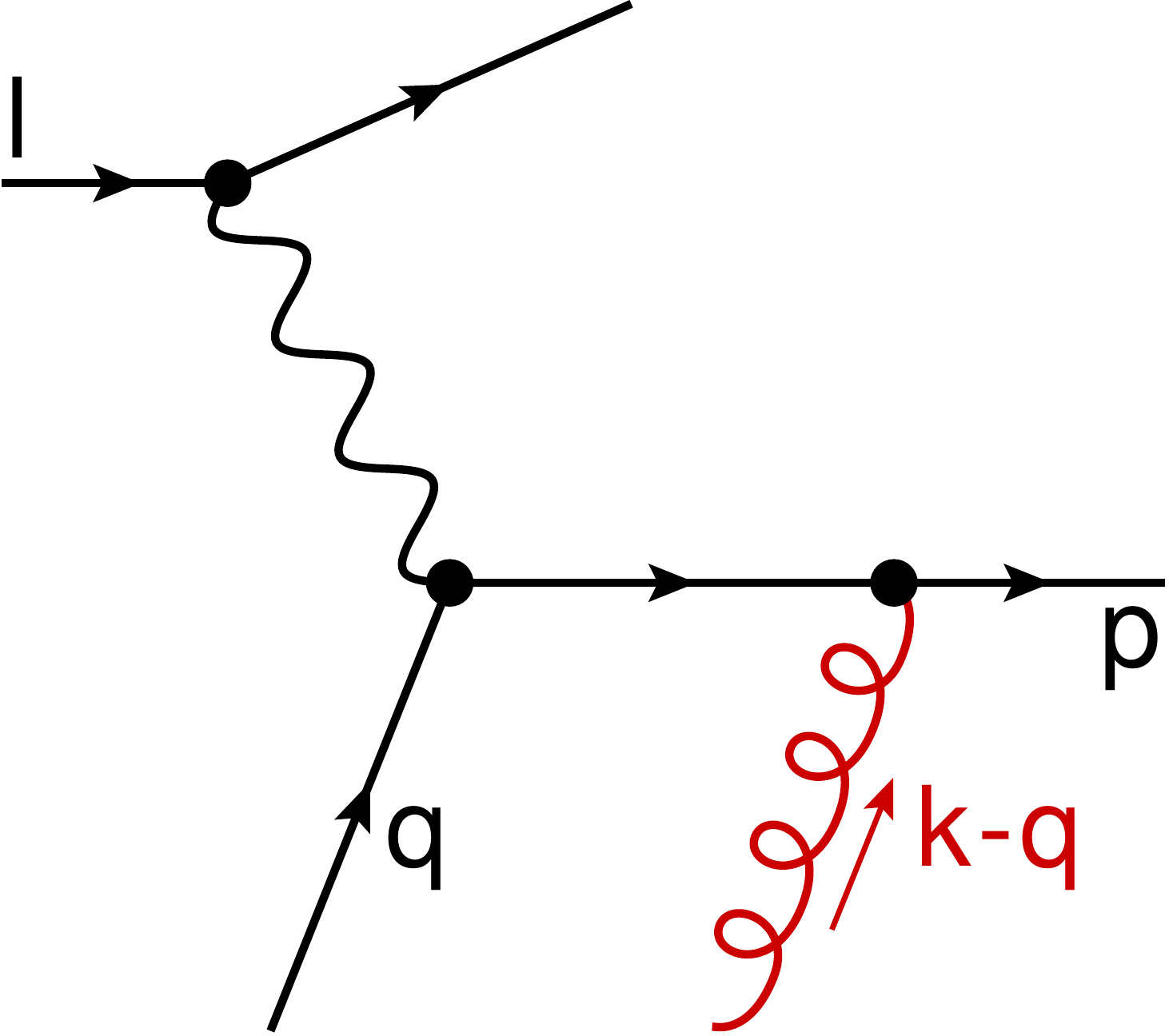}\label{fig:qg2qLO}}

\caption{Subfigure (a): Factorization (of the amplitude) in terms of quark-gluon-quark correlators for distributions. The square blob stands for the hard partonic scattering while the bottom oval blob represents the three-parton distribution functions, and the top oval blob represents a twist-2 fragmentation. This diagram describes dynamical twist-3 effects. Subfigure (b): Leading-order diagram for the hard partonic scattering.\label{fig:qg2qFIG}}
\end{figure*}

In order to obtain a LO result for our example process, one has to calculate the intrinsic, kinematical, and dynamical twist-3 contributions.  The diagram in Fig.~\ref{fig:q2q}, with the hard part shown in Fig.~\ref{fig:q2qLO}, yields the intrinsic and kinematical pieces. The dynamical parts are sketched in Figs.~\ref{fig:qg2q} and~\ref{fig:q2qg}, with their hard scatterings shown in Figs.~\ref{fig:qg2qLO} and~\ref{fig:q2qgLO}, respectively.
We refrain from giving detailed computational steps of these transverse spin observables in the twist-3 formalism and refer to the aforementioned papers.  We note that we have performed the calculation in both Feynman gauge and lightcone gauge and also explicitly checked that all the cross sections satisfy electromagnetic gauge invariance.  We will present our results in the following and will neglect quark masses throughout.

\subsection{SSA \boldmath{$A_{UTU}$} of a transversely polarized nucleon}

We start with the SSA for a transversely polarized nucleon and explicitly give results for the different twist-3 contributions (intrinsic, kinematical, and dynamical).  Even though we have performed the calculation in both Feynman gauge and lightcone gauge, the intermediate steps given here are for Feynman gauge.  We note that there are different approaches on how to carry out this computation (see, e.g., \cite{Eguchi:2006mc,Boer:1999si}), and the technique presented here follows more closely Ref.~\cite{Boer:1999si}. 

The calculation of the intrinsic twist-3 effects for the SSA of a transversely polarized nucleon is straightforward and can be obtained from Fig.~\ref{fig:q2qFIG}. At leading order we obtain the following result, 
\begin{eqnarray}
E_{h}\frac{d\sigma_{\,\mathrm{LO}}^{\mathrm{int.\, tw3}}(S_{N})}{d^{3}\vec{P}_{h}} & \!\!\!= \!\!\!& \frac{8\alpha_{\mathrm{em}}^{2}}{S}\sum_{q}e_{q}^{2}\int_{0}^{1}\frac{dx}{x}\,\int_{0}^{1}\frac{dz}{z^{3}}\,\delta(s+t+u)\times\nonumber \\
 &  &\hspace{-2.5cm} xM_{h}\,h_{1}^{q}(x)\, \frac{H^{q}(z)}{z}\,\Bigg[\frac{\epsilon^{lPP_{h}S_{N}}} {t^2}\,\left(\frac{\eta zs}{x}-1\right)\nonumber \\
 &  &\hspace{-0.3cm} +\left(\frac{T}{S}(l\cdot S_{N})+(P_{h}\cdot S_{N})\right)\,\frac{\eta_{\epsilon}z^{2}}{4x^{2}}\left(\frac{s^{2}u}{t^{2}}\right)\Bigg]\,,\label{eq:genTw3UPU}
\end{eqnarray}
where $\sum_{q}$ runs over all quarks and antiquarks. The partonic Mandelstam variables $s=xS$, $t=\frac{x}{z}T$, $u=\frac{1}{z}U$ were introduced in Eq.~(\ref{eq:genTw3UPU}) as well. We note that the parameters $\eta$ and $\eta_\epsilon$, which specify the lightcone vector $m^\mu$ in (\ref{LCVectors}), explicitly appear in the hard scattering. Hence, the intrinsic twist-3 contribution, considered on its own, depends on the choice of $m^\mu$ and is therefore frame-dependent. 

The kinematical twist-3 contribution to the nucleon SSA can be derived from Fig.~\ref{fig:q2qFIG} as well, and we obtain explicitly
\begin{eqnarray}
E_{h}\frac{d\sigma_{\mathrm{LO}}^{\mathrm{kin.\, tw3}}(S_{N})}{d^{3}\vec{P}_{h}} & \!\!\!=\!\!\! & \frac{8\alpha_{\mathrm{em}}^{2}}{S}\sum_{q}e_{q}^{2}\int_{0}^{1}\frac{dx}{x}\,\int_{0}^{1}\frac{dz}{z^{3}}\,\delta(s+t+u)\times\nonumber \\
 &  &\hspace{-2.8cm} \Bigg\{xM\,D_{1}^{q}(z)\,\Bigg[ \epsilon^{lPP_{h}S_{N}}\,f_{1T}^{\perp(1),q}(x)\,\left(\frac{(s-u)^{2}-tu}{t^{3}u}\right)\nonumber \\
 &  &\hspace{-1.7cm}+x\frac{\mathrm{d}}{\mathrm{d}x}\left(f_{1T}^{\perp(1),q}(x)\right)\left(\frac{s^{2}+u^{2}}{t^{2}}\right)\Bigg[\frac{\epsilon^{lPP_{h}S_{N}}} {s}\,\left(\frac{1}{u}-\frac{\chi z}{2x}\right)\nonumber \\
 &  &\hspace{-1cm}-\left(\frac{T}{S}(l\cdot S_{N})+(P_{h}\cdot S_{N})\right)\frac{\chi_{\epsilon}zs}{8x^{3}}\Bigg]\Bigg]\nonumber \\
 &  &\hspace{-2.7cm} +\,xM_{h}\, h_{1}^{q}(x)\,\Bigg[H_{1}^{\perp(1),q}(z)\,\Bigg[\frac{\epsilon^{lPP_{h}S_{N}}} {t^2}\,\left(\frac{4s}{t}+\frac{\eta zs}{x}\right)\nonumber\\
 &  &\hspace{-0.6cm} +\left(\frac{T}{S}(l\cdot S_{N})+(P_{h}\cdot S_{N})\right)\,\frac{\eta_{\epsilon}z^{2}}{4x^{2}}\,\left(\frac{s^2u}{t^{2}}\right)\Bigg]\nonumber \\
 &  & \hspace{-1.1cm}\,+z\frac{\mathrm{d}}{\mathrm{d}z}\left(H_{1}^{\perp(1),q}(z)\right)\,\Bigg[\frac{\epsilon^{lPP_{h}S_{N}}} {t^2}\left(2-\frac{\eta zs}{x}\right)\nonumber \\
 &  & \hspace{-0.6cm}-\left(\frac{T}{S}(l\cdot S_{N})+(P_{h}\cdot S_{N})\right)\frac{\eta_{\epsilon}z^{2}}{4x^{2}}\,\left(\frac{s^2u}{t^{2}}\right)\Bigg]\Bigg]\Bigg\}.\label{eq:kinTw3UPUmod}
\end{eqnarray}
Again we find an explicit dependence of the kinematical twist-3 contributions on the specific choice of the lightcone vectors $n^\mu$ and $m^\mu$.
\begin{figure*}[t]
\centering
\subfloat[]{\includegraphics[width=0.30\textwidth]{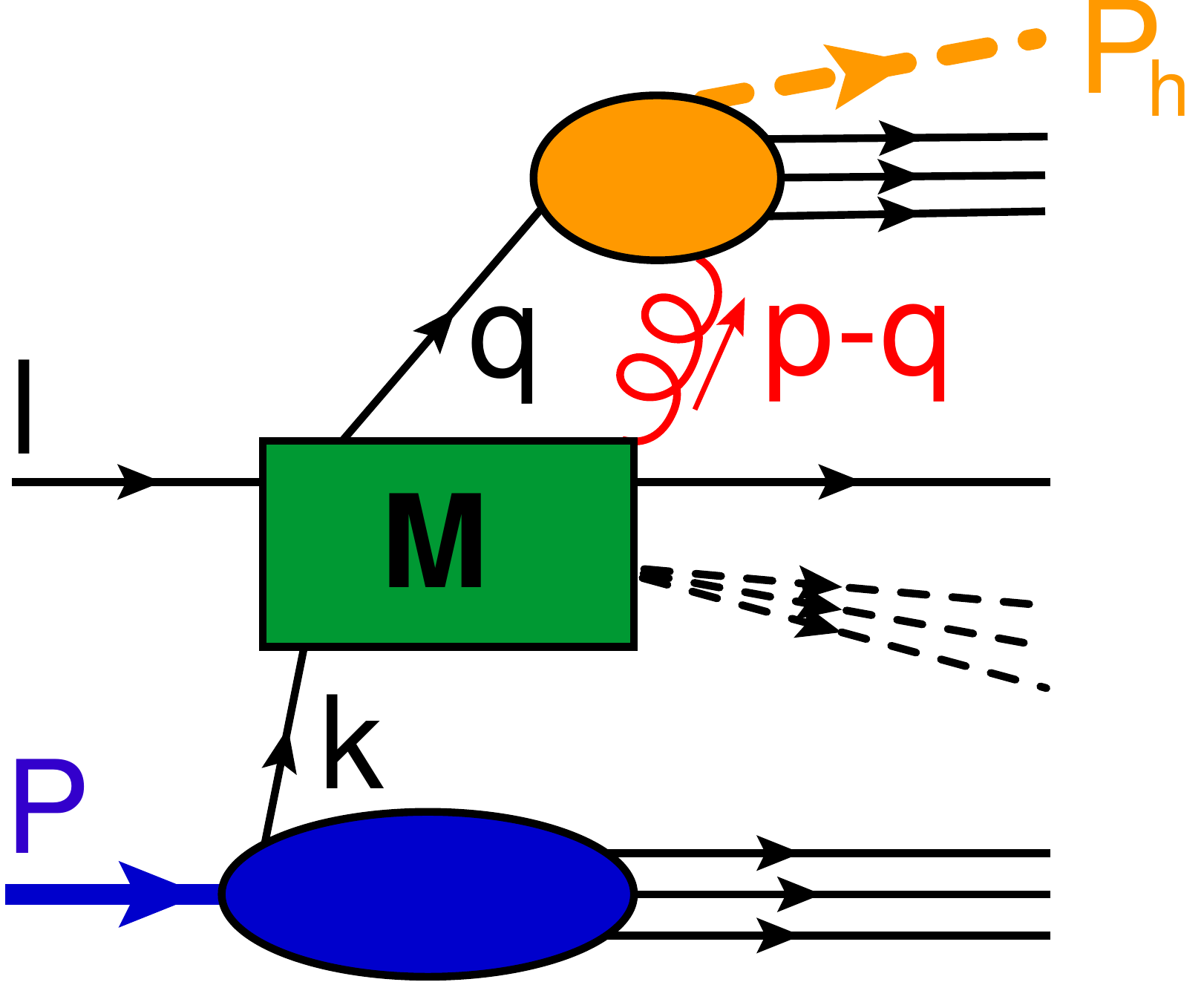}\label{fig:q2qg}}\hspace{1.5cm}
\subfloat[]{\includegraphics[width=0.30\textwidth]{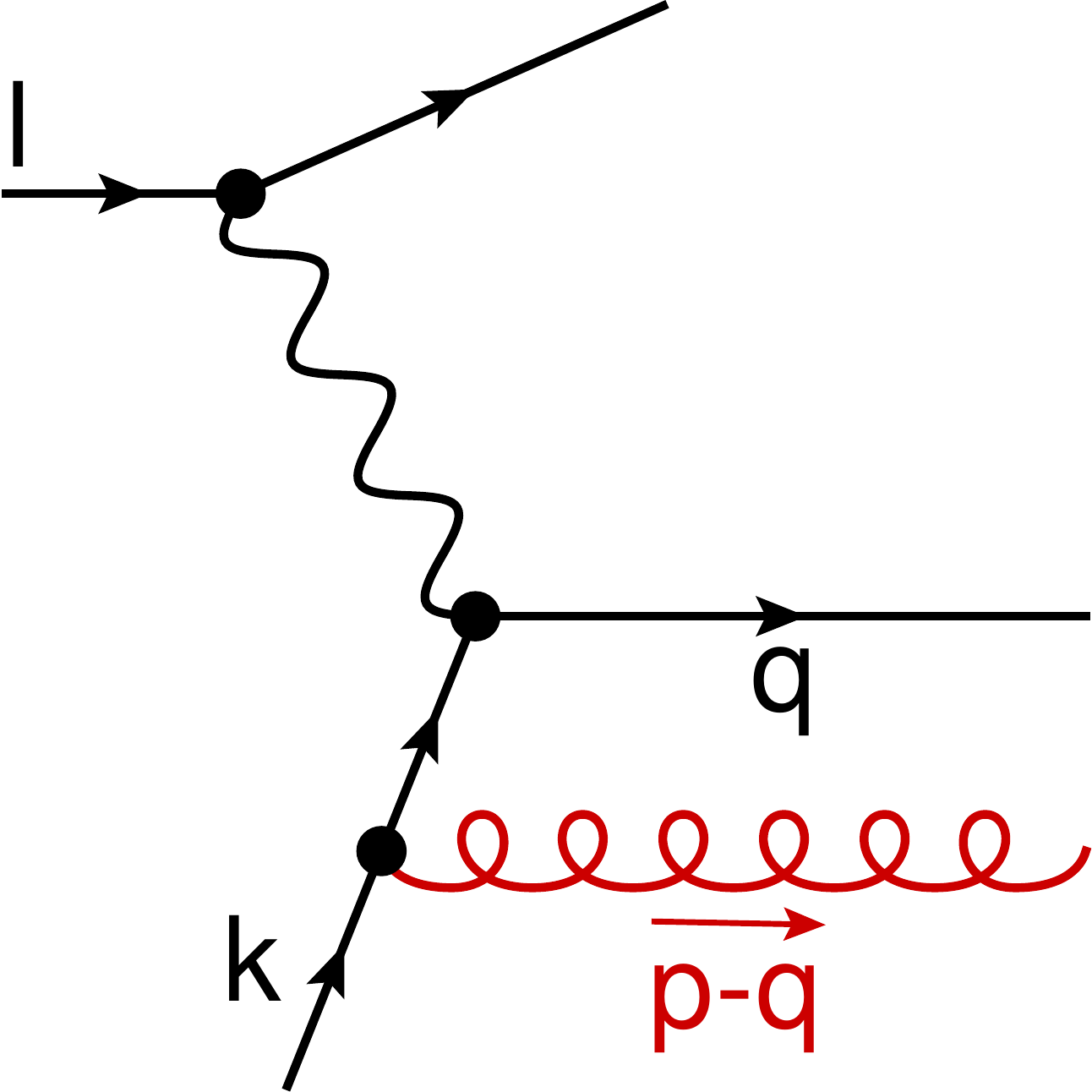}\label{fig:q2qgLO}}

\caption{Subfigure (a): Factorization (of the amplitude) in terms of quark-gluon-quark correlators for fragmentation. The square blob stands for the hard partonic scattering while the top oval blob represents the three-parton fragmentation functions, and the bottom oval blob represents a twist-2 parton distribution. This diagram describes dynamical twist-3 effects. Subfigure (b): Leading-order diagram for the hard partonic scattering.\label{fig:q2qgFIG}}
\end{figure*}
As a last step we have to calculate the dynamical twist-3 contributions from Figs.~\ref{fig:qg2qFIG} and \ref{fig:q2qgFIG}.
We find the result
\begin{eqnarray}
E_{h}\frac{d\sigma_{\mathrm{LO}}^{\mathrm{dyn.\, tw3}}(S_{N})}{d^{3}\vec{P}_{h}} & \!\!\!=\!\!\! & \frac{8\alpha_{\mathrm{em}}^{2}}{S}\sum_{q}e_{q}^{2}\int_{0}^{1}\frac{dx}{x}\,\int_{0}^{1}\frac{dz}{z^{3}}\,\delta(s+t+u)\times\nonumber \\
 &  &\hspace{-2.7cm} \Bigg\{xM\, D_{1}^{q}(z)\,\Bigg[\epsilon^{lPP_{h}S_{N}}\,\pi F_{FT}^{q}(x,x)\,\left(\frac{s-u}{t^{3}}\right)\nonumber \\
 &  &\hspace{-1.7cm}+\,\pi x\frac{\mathrm{d}}{\mathrm{d}x}\left(F_{FT}^{q}(x,x)\right)\left(\frac{s^{2}+u^{2}}{t^{2}}\right)\,\Bigg[\frac{\epsilon^{lPP_{h}S_{N}}} {s}\left(\frac{1}{t}+\frac{\chi z}{2x}\right)\nonumber \\
 &  & \hspace{-0.5cm}+\left(\frac{T}{S}(l\cdot S_{N})+(P_{h}\cdot S_{N})\right)\frac{\chi_{\epsilon}zs}{8x^{3}}\Bigg]\Bigg]\nonumber \\
  &  &\hspace{-2.5cm} +xM_{h}\, h_{1}^{q}(x)\,\Bigg[\epsilon^{lPP_{h}S_{N}}\left(2\int_{z}^{\infty}\frac{dz^{\prime}}{z^{\prime2}}\frac{\Im[\hat{H}_{FU}^{q}](z,z^{\prime})}{\frac{1}{z}-\frac{1}{z^{\prime}}}\right)\,\left(-\frac{s-u}{t^{3}}\right)\nonumber \\
 &  & \hspace{-2cm}+\,\left(\frac{2}{z}\int_{z}^{\infty}\frac{dz^{\prime}}{z^{\prime2}}\frac{\Im[\hat{H}_{FU}^{q}](z,z^{\prime})}{\left(\frac{1}{z}-\frac{1}{z^{\prime}}\right)^{2}}\right)\left(\frac{2s}{t^{2}}\right)\Bigg[\epsilon^{lPP_{h}S_{N}}\left(\frac{1}{t}+\frac{\eta z}{2x}\,\right)\nonumber \\
 &  & \hspace{-0.3cm}+\,\left(\frac{T}{S}(l\cdot S_{N})+(P_{h}\cdot S_{N})\right)\frac{\eta_{\epsilon}z^{2}su}{8x^{2}}\Bigg]\Bigg]\Bigg\},\label{eq:DynTw3UPU}
\end{eqnarray}
where we have used the identities $\left(\frac{\partial }{\partial x^{\prime}}F^q_{FT}(x,x')\right)\Big|_{x^{\prime}=x}\!=\!\frac{1}{2}\frac{\mathrm{d}}{\mathrm{d}x}F^q_{FT}(x,x)$
and $G_{FT}^{q}(x,x)=0$ from the symmetry properties discussed below Eq.~(\ref{eq:ParamDeltaF}), as well as the identity $\frac{\partial}{\partial (1/z^\prime)}\Re[\hat{H}^q_{FU}(z,z^\prime)]|_{z^\prime=z}=0$ derived in Appendix \ref{App:AppendixA}. 

In order to find the full LO result, we have to add Eqs.~(\ref{eq:genTw3UPU}), (\ref{eq:kinTw3UPUmod}), and (\ref{eq:DynTw3UPU}). In doing so, one sees the essential role that LIRs (and other operator relations involving gluonic pole matrix elements) play in ensuring a frame-independent result.  For example, adding the coefficients of $\chi$ and $\chi_\epsilon$ of the derivative terms of the functions $f_{1T}^{\perp (1),q}(x)$ in Eq.~(\ref{eq:kinTw3UPUmod}) and of $F^q_{FT}(x,x)$ in Eq.~(\ref{eq:DynTw3UPU}) leads to the term
\begin{align}
&\frac{\mathrm{d}}{\mathrm{d}x}\left(\pi F^q_{FT}(x,x)-f_{1T}^{\perp (1),q}(x)\right)\,D_1^q(z)\,\left(\frac{s^2+u^2}{st^2}\right) \times&\nonumber\\
&\hspace{0.5cm}\left(\frac{\chi z}{2x }\epsilon^{lPP_{h}S_{N}}+\frac{\chi_\epsilon z s^2}{8x^3}\,\left(\frac{T}{S}(l\cdot S_{N})+(P_{h}\cdot S_{N})\right)\right).&
\end{align}
We see that this term vanishes (as it must) due to the well-known relation between the (gluonic pole) Qiu-Sterman matrix element $F_{FT}(x,x)$ \cite{Qiu:1991pp,Qiu:1991wg,Qiu:1998ia,Kouvaris:2006zy} and the transverse-momentum moment of the Sivers function~\cite{Sivers:1989cc},
\begin{equation}
\pi F^q_{FT}(x,x)=f_{1T}^{\perp (1),q}(x)\,, \label{GluonicPole}
\end{equation}
which has been derived on the operator level in Ref.~\cite{Boer:2003cm}. We note that (\ref{GluonicPole}) holds for a Wilson line as defined in Eq.~(\ref{PhiTMD}) which is typically relevant in a TMD description of semi-inclusive deep-inelastic scattering (SIDIS) processes (cf. Ref.~\cite{Bacchetta:2006tn}). 
(We note that, in principle, we could reverse the argument and ``re-derive'' (\ref{GluonicPole}) by demanding that physical observables be independent of the choice of lightcone vectors.)

Let us now look at the term proportional to the parameters $\eta$ and $\eta_\epsilon$ when adding the intrinsic, kinematical, and dynamical twist-3 contributions (\ref{eq:genTw3UPU}), (\ref{eq:kinTw3UPUmod}), and (\ref{eq:DynTw3UPU}). 
In this sum we find the term:
\begin{align}
&h_1^q(x)\left(\frac{H^q(z)}{z}+\left(1-z\frac{\mathrm{d}}{\mathrm{d}z}\right)H_1^{\perp (1),q}(z)+\frac{2}{z}\int_{z}^{\infty}\frac{dz^{\prime}}{z^{\prime2}}\frac{\Im[\hat{H}_{FU}^{q}](z,z^{\prime})}{\left(\frac{1}{z}-\frac{1}{z^{\prime}}\right)^{2}}\right)&\nonumber\\
&\times\left(\frac{\eta z}{2x}\epsilon^{lPP_{h}S_{N}}+\frac{\eta_\epsilon z^2 s u}{8x^2}\,\left(\frac{T}{S}(l\cdot S_{N})+(P_{h}\cdot S_{N})\right)\right)\!\left(\frac{2s}{t^2}\right).
\end{align}
We see that this piece vanishes due to the LIR we have derived in Eq.~(\ref{LIRH}). 
(As before, we note that requiring physical observables to be independent of the choice of lightcone vectors leads to a ``re-derivation'' of the relation (\ref{LIRH}).)
Though in our example we have considered only the LO result, we conjecture that the cancelation of the parameters $\chi$, $\eta$, $\chi_\epsilon$, $\eta_\epsilon$ should hold to all orders by means of the LIRs derived in the previous section.

By applying the relations (\ref{GluonicPole}), (\ref{LIRH}), as well as the e.o.m.~relation (\ref{EoMH}), to the intrinsic, kinematical, and dynamical twist-3 contributions (\ref{eq:genTw3UPU}), (\ref{eq:kinTw3UPUmod}), and (\ref{eq:DynTw3UPU}), we can achieve an enormous simplification and obtain a compact result for the nucleon SSA,
\begin{eqnarray}
E_{h}\frac{d\sigma_{\,\mathrm{LO}}(S_{N})}{d^{3}\vec{P}_{h}} & = & \frac{8\alpha_{\mathrm{em}}^{2}}{S}\,\sum_{q}e_{q}^{2}\int_{0}^{1}dx\,\int_{0}^{1}\frac{dz}{z^{3}}\,\delta(s+t+u)\nonumber \\
 &  &\hspace{-2.3cm} \times \, \epsilon^{lPP_{h}S_{N}}\,\Bigg[\pi M\,\left(1-x\frac{\mathrm{d}}{\mathrm{d}x}\right)F_{FT}^{q}(x,x)\, D_{1}^{q}(z)\,\left(\frac{s^{2}+u^{2}}{t^{3}u}\right)\nonumber \\
 &  & \hspace{-2.3cm}+M_{h}\, h_{1}^{q}(x)\,\Bigg(\frac{H^{q}(z)}{z}\,\left(\frac{2(u-s)}{t^{3}}\right) \nonumber\\
& &+\,\left(1-z\frac{\mathrm{d}}{\mathrm{d}z}\right)H_{1}^{\perp(1),q}(z)\,\left(\frac{2u}{t^{3}}\right)\Bigg)\Bigg]\,. \label{eq:UPUFin}
\end{eqnarray}
This agrees with Ref.~\cite{Gamberg:2014eia} after we apply~(\ref{LIRH}) to their result.  We remind the reader that Ref.~\cite{Eguchi:2006mc} provides another Feynman-gauge formalism for calculating the pole contributions from the twist-3 distribution correlator,
which directly leads to the $F_{FT}(x,x)$-contribution in Eq.~(\ref{eq:UPUFin}) without
recourse to the intermediate steps given in (\ref{eq:kinTw3UPUmod}) and (\ref{eq:DynTw3UPU}).  

The formula (\ref{eq:UPUFin}) contains two twist-3 fragmentation functions, $H$ and $H_1^{\perp (1)}$. By means of Eqs.~(\ref{H1perp1}) and (\ref{H}), we can express these functions in terms of a single twist-3 quark-gluon-quark fragmentation function $\Im[\hat{H}_{FU}]$. The nucleon SSA then takes the final form,
\begin{eqnarray}
E_{h}\frac{d\sigma_{\,\mathrm{LO}}(S_{N})}{d^{3}\vec{P}_{h}} & = & \frac{8\alpha_{\mathrm{em}}^{2}}{S}\,\sum_{q}e_{q}^{2}\int_{0}^{1}\frac{dx}{x^{2}}\,\int_{0}^{1}\frac{dz}{z}\,\delta(s+t+u)\nonumber \\
 &  &\hspace{-2cm}\times\, \frac{\epsilon^{lPP_{h}S_{N}}}{T^{2}}\, \Bigg[M\,\left(1-x\frac{\mathrm{d}}{\mathrm{d}x}\right)\pi\, F_{FT}^{q}(x,x)\, D_{1}^{q}(z)\,\left(\frac{s^{2}+u^{2}}{tu}\right)\nonumber \\
 &  & \hspace{-2cm}+M_{h}\, h_{1}^{q}(x)\,\Bigg(-\frac{4}{z}\int_{z}^{1}dz_{1}\int_{z_{1}}^{\infty}\frac{dz_{2}}{z_{2}^{2}}\frac{\Im[\hat{H}_{FU}^{q}(z_{1},z_{2})]}{\left(\frac{1}{z_{1}}-\frac{1}{z_{2}}\right)^{2}}\times\nonumber \\
 &  &\hspace{-2.2cm} \Bigg[\frac{2s}{t}\left(\frac{2}{z_{1}}-\frac{1}{z_{2}}\right)\left(1+\frac{1}{2z_{1}}\delta\left(\frac{1}{z}-\frac{1}{z_{1}}\right)\right)-\frac{s-u}{z_{1}^{2}t}\delta\left(\frac{1}{z}-\frac{1}{z_{1}}\right)\Bigg]\Bigg)\Bigg].\nonumber\\ \label{eq:UPUFinqgq4}
\end{eqnarray}
We see that the nucleon SSA can be expressed in pQCD solely in terms of three-parton twist-3 correlation functions (along with twist-2 distribution/fragmentation functions like $h_1$ and $D_1$).  Since we find the same behavior for the other transverse spin observables (see below), we conclude that these three-parton correlation functions are the key ingredients for the description of twist-3 observables whereas intrinsic and kinematical twist-3 functions may be viewed as auxiliary functions.

\subsection{DSA \boldmath{$A_{LTU}$} of a longitudinally polarized lepton and a transversely polarized nucleon}
We proceed in a similar fashion with the DSA of a longitudinally polarized lepton and a transversely polarized nucleon. Upon adding the intrinsic, kinematical, and dynamical twist-3 effects we find again that any dependence on the parameters $\chi$, $\eta$, $\chi_\epsilon$, and $\eta_\epsilon$ cancels when the LIR (\ref{LIRgT}) is applied. By means of the e.o.m.~relations (\ref{EoMgT}) and (\ref{EoME}) we find a compact result that was already given in Ref.~\cite{Kanazawa:2014tda},
\begin{eqnarray}
E_h\frac{d\Delta\sigma_{\mathrm{LO}}(S_{N})} {d^3\vec{P}_h} & \equiv &\frac{1}{2}\left(E_{h}\frac{d\sigma_{\mathrm{LO}}^{\lambda_{\ell}=+1}(S_{N})}{d^{3}\vec{P}_{h}}-E_{h}\frac{d\sigma_{\mathrm{LO}}^{\lambda_{\ell}=-1}(S_{N})}{d^{3}\vec{P}_{h}}\right) \nonumber\\
&  &\hspace{-2.1cm}=\frac{8\alpha_{\mathrm{em}}^{2}}{S}\sum_{q}e_{q}^{2}\int_{0}^{1}\frac{dx}{x}\,\int_{0}^{1}\frac{dz}{z^3}\,\delta(s+t+u)\,\times \nonumber \\
&&\hspace{-1.9cm}\Bigg[\,\left(\frac{z^2M\,(l\cdot S_{N})}{-2T}\right)\,g_{1}^{q}(x)\,D_{1}^{q}(z)\,\left(\frac{s-u}{s}\right)\nonumber\\
 &  & \hspace{-1.9cm}+\left(\frac{T}{S}\,(l\cdot S_{N})+(P_{h}\cdot S_{N})\right)\, \Bigg[M\,\Bigg(x\, g_{1}^{q}(x)\left(\frac{s-u}{2t^{2}}\right)\nonumber \\
 &  &\hspace{-1.9cm} +x\, g_{T}^{q}(x)\,\left(\frac{-s}{t^{2}}\right)+\left(1-x\frac{\mathrm{d}}{\mathrm{d}x}\right)g_{1T}^{(1),q}(x)\,\left(\frac{s(s-u)}{2t^{2}u}\right)\Bigg)\, D_{1}^{q}(z)\nonumber \\
 &  &\hspace{-1.9cm} +M_{h}\, h_{1}^{q}(x)\, \frac{E^{q}(z)}{z}\left(\frac{-s}{t^{2}}\right)\Bigg]\,\Bigg].\label{eq:LPUFin}
\end{eqnarray}
Note that the first term of the integrand in this equation is the twist-2 DSA for a longitudinally polarized nucleon that is generated by the twist-2 helicity distribution $g_1$.
The DSA for a transversely polarized nucleon is generated by $g_1$, the twist-3 distribution functions $g_T$ and $g_{1T}^{(1)}$, and the twist-3 fragmentation function $E$. Again, by applying the relations (\ref{EoME}), (\ref{g1T}), and (\ref{gT}) we can express this double-spin observable in terms of three-parton correlation functions,
\begin{eqnarray}
E_h\frac{d\Delta\sigma_{\mathrm{LO}}(S_{N})} {d^3\vec{P}_h} & = & \frac{8\alpha_{\mathrm{em}}^{2}}{S}\sum_{q}e_{q}^{2}\int_{0}^{1}\frac{dx}{x}\,\int_{0}^{1}\frac{dz}{z^3}\,\delta(s+t+u)\,\times\nonumber \\
&&\hspace{-1.5cm}\Bigg[\,\left(\frac{z^2M\,(l\cdot S_{N})}{-2T}\right)\,g_{1}^{q}(x)\, D_{1}^{q}(z)\,\left(\frac{s-u}{s}\right) \nonumber\\
 &  &\hspace{-1.5cm} +\left(\frac{T}{S}\,(l\cdot S_{N})+(P_{h}\cdot S_{N})\right)\times\nonumber\\
&&\hspace{-1.5cm} \,\left[xM\left\{ g_1^q(x)\left(\frac{u-s} {2tu}\right)-\frac{s} {t^2}\int_x^1\!dx'\,\frac{g_1(x')} {x'}\right.\right.\nonumber\\
&&\hspace{-1.5cm} -\,\frac{s} {t^2}\int_x^1\!\frac{dx_1} {x_1^2}\,\mathcal{P}\!\int_{-1}^1\!dx_2\left[\frac{1-x_1\delta(x_1-x)} {x_1-x_2}\,F_{FT}^q(x_1,x_2)\right.\nonumber\\
&&\hspace{-1cm}\left.-\,\frac{3x_1-x_2-x_1(x_1-x_2)\delta(x_1-x)} {(x_1-x_2)^2}\,G^q_{FT}(x_1,x_2)\right]\nonumber\\
&&\hspace{-1.5cm}\left.+\,\frac{s(s-u)} {2xt^2u}\,\mathcal{P}\!\int_{-1}^1\!dx'\left[\frac{1} {x-x'}\,F^q_{FT}(x,x')\right.\right.\nonumber\\
&&\hspace{1.3cm}\left.\left.-\,\frac{3x-x'} {(x-x')^2}\,G^q_{FT}(x,x')\right]\right\}\,D_1^q(z)\nonumber\\
 &  & \hspace{-1.5cm}+\,M_{h}\frac{2s} {t^2}\, h_{1}^{q}(x)\,\int_{z}^{\infty}\frac{dz^{\prime}}{z^{\prime2}}\frac{\Re[\hat{H}_{FU}^{q}(z,z^{\prime})]}{\frac{1}{z}-\frac{1}{z^{\prime}}}\Bigg]\,\Bigg]\,.\label{eq:LPUFinqgq}
\end{eqnarray}
Note that the contribution in~(\ref{eq:LPUFinqgq}) that is given by $g_1$ may be considered the WW approximation to the DSA $A_{LTU}$ upon the assumption that the three-parton correlation functions are small.

\subsection{SSA \boldmath{$A_{UUT}$} of a transversely polarized hadron}
We now consider the SSA of a transversely polarized (outgoing) hadron. We again follow the same steps as for the other observables, i.e., add the intrinsic, kinematical, and dynamical twist-3 contributions. 
Similar to the nucleon SSA, the coefficient of the parameters $\chi$ and $\chi_\epsilon$ vanishes due to a relation involving the (gluonic pole) matrix element $H_{FU}(x,x)$ and the transverse-momentum moment $h_1^{\perp (1)}(x)$ of the Boer-Mulders function~\cite{Boer:1997nt}.  The identity, first derived on the operator level in~\cite{Boer:2003cm} for a Wilson line as defined in Eq.~(\ref{PhiTMD}), reads
\begin{equation}
\pi H_{FU}^q(x,x)=h_1^{\perp (1),q}(x)\,.\label{GluonicPoleBM}
\end{equation}
Likewise, the coefficient of $\eta$ and $\eta_\epsilon$ vanishes due to the LIR (\ref{LIRDT}). In addition to (\ref{GluonicPoleBM}) and (\ref{LIRDT}), one may also use the e.o.m.~relation (\ref{EoMDT}) in order to obtain a compact result, which is in agreement with the one in Ref.~\cite{Kanazawa:2015jxa} after we apply (\ref{LIRDT}) to their calculation,
\begin{eqnarray}
E_{h}\frac{d\sigma_{\mathrm{LO}}(S_{h})}{d^{3}\vec{P}_{h}} & = & \frac{8\alpha_{\mathrm{em}}^{2}}{S}\,\sum_{q}e_{q}^{2}\int_{0}^{1}dx\,\int_{0}^{1}\frac{dz}{z^{3}}\,\delta(s+t+u)\times\nonumber \\
 &  & \hspace{-1.7cm}\epsilon^{lPP_{h}S_{h}}\,\Bigg(M\,\Bigg[\pi\,x\frac{\mathrm{d}H_{FU}^{q}(x,x)}{\mathrm{d}x}\,\left(\frac{2s}{t^{3}}\right)\Bigg]\, H_{1}^{q}(z)\label{eq:UUPFin} \\
 &  & \hspace{-2.2cm}+M_{h}\, f_{1}^{q}(x)\,\Bigg[\frac{D_{T}^{q}(z)}{z}\left(\frac{-2(s-u)}{t^{3}}\right)+z\frac{\mathrm{d}D_{1T}^{\perp(1),q}(z)}{\mathrm{d}z}\,\left(\frac{s^{2}+u^{2}}{st^{3}}\right)\Bigg]\Bigg)\,.\nonumber
\end{eqnarray}
We may insert the expressions (\ref{D1Tperp1}) and (\ref{DT}) into (\ref{eq:UUPFin}) in order to write the observable in terms of three-parton correlation functions,
\begin{eqnarray}
E_{h}\frac{d\sigma_{\mathrm{LO}}(S_{h})}{d^{3}\vec{P}_{h}} & = & \frac{8\alpha_{\mathrm{em}}^{2}}{S}\,\sum_{q}e_{q}^{2}\int_{0}^{1}dx\,\int_{0}^{1}\frac{dz}{z^{3}}\,\delta(s+t+u)\times\nonumber \\
 &  &\hspace{-1.7cm} \epsilon^{lPP_{h}S_{h}}\,\Bigg(M\,\Bigg[\pi\,x\frac{\mathrm{d}H_{FU}^{q}(x,x)}{\mathrm{d}x}\,\left(\frac{2s}{t^{3}}\right)\Bigg]\, H_{1}^{q}(z)\nonumber \\
 &  & \hspace{-1.7cm}+M_{h}\, f_{1}^{q}(x)\,\Bigg[\frac{-2(s-u)}{t^{3}}\int_{z}^{1}\frac{dz_{1}}{z_{1}}\int_{z_{1}}^{\infty}\frac{dz_{2}}{z_{2}^{2}}\times\nonumber \\
 &  &\hspace{-1.7cm} \Bigg[\frac{\left(\frac{3}{z_{1}}-\frac{1}{z_{2}}+z\left(\frac{1}{z_{1}}-\frac{1}{z_{2}}\right)\delta(z_{1}-z)\right)\Im[\hat{D}_{FT}^{q}(z_{1},z_{2})]}{\left(\frac{1}{z_{1}}-\frac{1}{z_{2}}\right)^{2}}\nonumber \\
&  &\hspace{-1.7cm}-\frac{\left(1+z\,\delta(z_{1}-z)\right)\Im[\hat{G}_{FT}^{q}(z_{1},z_{2})]}{\left(\frac{1}{z_{1}}-\frac{1}{z_{2}}\right)}\Bigg]\nonumber \\
 &  &\hspace{-1.7cm} +\frac{s^{2}+u^{2}}{st^{3}}\int_{z}^{\infty}\frac{dz^{\prime}}{z^{\prime2}}\Bigg[\frac{\left(\frac{3}{z}-\frac{1}{z^{\prime}}\right)\Im[\hat{D}_{FT}^{q}(z,z^{\prime})]}{\left(\frac{1}{z}-\frac{1}{z^{\prime}}\right)^{2}}\nonumber\\
&  &\hspace{-1.7cm} -\frac{\Im[\hat{G}_{FT}^{q}(z,z^{\prime})]}{\left(\frac{1}{z}-\frac{1}{z^{\prime}}\right)}\Bigg]\,\Bigg]\Bigg)\,.\label{eq:UUPFinqgq4}
\end{eqnarray}

\subsection{DSA \boldmath{$A_{LUT}$} of a longitudinally polarized lepton and a transversely polarized hadron}

To complete this list of example observables we also calculate the DSA of a longitudinally polarized lepton and a transversely polarized (outgoing) hadron. As before, we add all twist-3 contributions and see that the coefficients of the parameters $\chi$, $\chi_\epsilon$, $\eta$, $\eta_\epsilon$ vanish due to the LIR (\ref{LIRGT}). Then, using the e.o.m.~relations (\ref{EoMe}) and (\ref{EoMGT}), we obtain the following result,
\begin{eqnarray}
E_h\frac{d\Delta\sigma_{\mathrm{LO}}(S_{h})} {d^3\vec{P}_h}&\!\!\!=\!\!\!&\frac{8\alpha_{\mathrm{em}}^{2}}{S}\,\sum_{q}e_{q}^{2}\int_{0}^{1}dx\,\int_{0}^{1}\frac{dz}{z^{3}}\,\delta(s+t+u)\,\times\nonumber \\
&&\hspace{-1.5cm}\Bigg\{ \left(\frac{zM_{h}(l\cdot S_{h})}{-2xU}\right)\,\, f_{1}^{q}(x)\, G_{1}^{q}(z)\,\left(\frac{u-s}{t}\right)\nonumber\\
 &  & \hspace{-1.5cm}+\left(-\frac{T}{U}(l\cdot S_{h})+(P\cdot S_{h})\right)\, \Bigg[xM\, e^{q}(x)\, zH_{1}^{q}(z)\,\left(\frac{u}{t^{2}}\right)\nonumber \\
 &  &\hspace{-1.5cm} +zM_{h}\, f_{1}^{q}(x)\,\Bigg[\frac{G_{1}^{q}(z)}{z}\,\left(\frac{s-u}{2t^{2}}\right)+\frac{G_{T}^{q}(z)}{z}\,\left(\frac{u}{t^{2}}\right)\nonumber\\
& & \hspace{-1.5cm}+z\frac{\text{\ensuremath{\mathrm{d}}}}{\mathrm{d}z}\left(G_{1T}^{(1),q}(z)\right)\,\left(\frac{u(s-u)}{2st^{2}}\right)\Bigg]\Bigg]\Bigg\}\,.\label{eq:LUPFin}
\end{eqnarray}
Like in Eq.~(\ref{eq:LPUFin}), we find in~(\ref{eq:LUPFin}) contributions for a longitudinally polarized hadron (terms proportional to $G_1$), whereas the rest describes transverse polarization.
Using (\ref{G1T1}) and (\ref{GT}) to express the intrinsic and kinematical twist-3 functions in terms of dynamical three-parton functions, we find
\begin{eqnarray}
E_h\frac{d\Delta\sigma_{\mathrm{LO}}(S_{h})} {d^3\vec{P}_h} & = & \frac{8\alpha_{\mathrm{em}}^{2}}{S}\,\sum_{q}e_{q}^{2}\int_{0}^{1}dx\,\int_{0}^{1}\frac{dz}{z^{3}}\,\delta(s+t+u)\, \times\nonumber\\
& & \hspace{-1.7cm}\Bigg\{ \left(\frac{zM_{h}(l\cdot S_{h})}{-2xU}\right)\, f_{1}^{q}(x)\, G_{1}^{q}(z)\,\left(\frac{u-s}{t}\right)\nonumber \\
 &  & \hspace{-1.7cm}+\left(-\frac{T}{U}(l\cdot S_{h})+(P\cdot S_{h})\right)\times\nonumber \\
 &  &\hspace{-1.7cm} \Bigg[-2M\,\left(\mathcal{P}\int_{-1}^{1}dx^{\prime}\frac{H_{FU}^{q}(x,x^{\prime})}{x-x^{\prime}}\right)\, z\, H_{1}^{q}(z)\,\left(\frac{u}{t^{2}}\right)\nonumber \\
 &  &\hspace{-1.7cm} +zM_{h}\, f_{1}^{q}(x)\,\Bigg[\frac{G_{1}^{q}(z)}{z}\,\left(\frac{u-s}{2st}\right)-\frac{u}{t^{2}}\int_{z}^1dz'\frac{G_{1}^{q}(z')}{z'^{2}}\nonumber \\
 &  &\hspace{-1.9cm} -\frac{u}{t^{2}}\int_{z}^{1}\frac{dz_{1}}{z_{1}^{2}}\int_{z_{1}}^{\infty}\frac{dz_{2}}{z_{2}^{2}}\left[\frac{\left(z_{1}+\delta(\tfrac{1}{z}-\tfrac{1}{z_{1}})\right)\,}{\frac{1}{z_{1}}-\frac{1}{z_{2}}}\Re[\hat{D}_{FT}^{q}(z_{1},z_{2})]\right.\nonumber \\
 & & \hspace{-1cm}-\left.\left(\frac{3-\frac{z_{1}}{z_{2}}+(\frac{1} {z_1}-\frac{1} {z_2})\,\delta(\tfrac{1}{z}-\tfrac{1}{z_{1}})}{\left(\frac{1}{z_{1}}-\frac{1}{z_{2}}\right)^{2}}\right)\Re[\hat{G}_{FT}^{q}(z_{1},z_{2})]\right]\nonumber \\
 &  &\hspace{-1.7cm} +\frac{u(s-u)}{2st^{2}}\int_{z}^{\infty}\frac{dz^{\prime}}{z^{\prime2}}\left(\frac{1}{\frac{1}{z}-\frac{1}{z^{\prime}}}\Re[\hat{D}_{FT}^{q}(z,z^{\prime})]\right. \nonumber\\
 &&\left.\left.\left.\left.-\,\frac{\frac{3} {z}-\frac{1}{z^{\prime}}}{\left(\frac{1}{z}-\frac{1}{z^{\prime}}\right)^{2}}\Re[\hat{G}_{FT}^{q}(z,z^{\prime})]\right)\right]\right]\right\} \,.\label{eq:LUPFinqgq4}
\end{eqnarray}

\section{Conclusions\label{s:concl}}
In this paper we have addressed the Lorentz invariance of twist-3 observables.  We derived so-called Lorentz invariance relations among twist-3 correlators based on the QCD e.o.m.~and identities among non-local operators, where, in particular, Eqs.~(\ref{LIRH}), (\ref{LIRDT}), (\ref{LIRGT}), (\ref{LIRHL}) for fragmentation functions are novel formulae from this work.  Moreover, we showed how intrinsic and kinematical twist-3 functions can be written in terms of dynamical (three-parton) functions --- see Eqs.~(\ref{g1T})--(\ref{GT}).  
Thus, twist-3 observables can be expressed exclusively through quark-gluon-quark (and tri-gluon) correlations.

In addition, we investigated why twist-3 observables are not manifestly frame-independent, i.e., why one needs LIRs to demonstrate their Lorentz invariance.  
We found that the frame-dependence enters through the lightcone vectors $n^\mu,\,m^\mu$ that are ``adjoint" vectors to the nucleon/hadron momenta.
Since twist-3 effects contain ``transverse'' quantities (e.g., a transverse spin component) that depend on these vectors, this initially leads to results that depend on the specific choice of $n^\mu,\,m^\mu$, which corresponds to the choice of frame. 
However, the use of LIRs subsequently eliminates this dependence, and one in the end has Lorentz invariant observables.  

Finally, we demonstrated this feature explicitly by calculating asymmetries in $\ell\,N\to h\,X$ for various lepton/nucleon/hadron polarizations:~$A_{UTU}$, $A_{LTU}$, $A_{UUT}$, $A_{LUT}$, with the last one a new result from this work.  
All results were checked in two color gauges (Feynman and lightcone) and shown to satisfy electromagnetic gauge invariance.  
Thus, we have further established the theoretical validity of the collinear twist-3 formalism used in the computation of both distribution and fragmentation effects for SSAs and DSAs in lepton-nucleon and proton-proton collisions.

\begin{acknowledgments}

K.K. thanks S.~Yoshida for useful discussions during his stay at Niigata
 University. This work has been supported by the National Science Foundation under Contract No.~PHY-1516088 (K.K. and A.M.), the Grant-in-Aid for Scientific Research from the Japanese Society of Promotion of Science under Contract No.~26287040 (Y.K.), the RIKEN BNL Research Center (D.P.), and the �Bundesministerium f\"{u}r Bildung und Forschung� (BMBF) grant 05P12VTCTG (M.S.).
\end{acknowledgments}

\appendix
\section{Proof that \boldmath{$\frac{\partial}{\partial (1/z^\prime)}\Delta_{F}^{\rho}(z,z^\prime)\Big|_{z^\prime = z}=0$}} \label{App:AppendixA}
In this Appendix we will prove that $\frac{\partial}{\partial (1/z^\prime)}\Delta_{F}^{\rho}(z,z^\prime)\Big|_{z^\prime = z}=0$, where $\Delta_{F}^{\rho}(z,z^\prime)$ is given in Eq.~(\ref{DeltaFzz'}), which we write here again for completeness:
\begin{align}
&\Delta^{q,\rho}_{F,ij}(z,z^\prime) = \frac{1}{N_c}\sumint  \int_{-\infty}^\infty\frac{d\lambda}{2\pi}\int_{-\infty}^\infty\frac{d\mu}{2\pi}\,\mathrm{e}^{i\frac{\lambda}{z^\prime}+i(\frac{1}{z}-\frac{1}{z^\prime})\mu}\label{DeltaFzz'Hc}\nonumber \\
&\hspace{0.75cm}\times\langle 0|\,ig\,m_\eta F^{\eta \rho}(\mu m)\,q_i(\lambda m)|P_hS_h\,;\,X\rangle \langle P_hS_h\,;\,X|\bar{q}_j(0)\,|0\rangle\,.
\end{align}
(We have ignored the Wilson lines for brevity as they will not affect our proof.)  The arguments follow closely those of Ref.~\cite{Meissner:2008yf}.  We start by taking the derivative of (\ref{DeltaFzz'Hc}) w.r.t. $1/z^\prime$, inserting a complete set of states in between $F^{\eta \rho}(\mu m)$ and $q_i(\lambda m)$, shifting the fields to a spacetime point of 0, and then setting $z^\prime = z$.  This gives
\begin{align}
&\frac{\partial \Delta^{q,\rho}_{F,ij}(z,z^\prime)} {\partial (1/z^\prime)}\Bigg |_{z=z^\prime} \!\!= \frac{1}{N_c}\sumintXY  \int_{-\infty}^\infty\frac{d\lambda}{2\pi}\int_{-\infty}^\infty\frac{d\mu}{2\pi}\,i(\lambda-\mu)\nonumber\\
&\hspace{1cm}\times\mathrm{e}^{i\lambda(\frac{1} {z}-1+(P_Y-P_X)\cdot m)}e^{-i\mu P_Y\cdot m}\,\langle 0|\,ig\,m_\eta F^{\eta \rho}(0)|Y\rangle\nonumber \\
&\hspace{1cm}\times\langle Y|\,q_i(0)|P_hS_h\,;\,X\rangle \langle P_hS_h\,;\,X|\bar{q}_j(0)\,|0\rangle \nonumber\\[0.3cm]
& = \frac{1}{N_c}\sumintXY  \int_{-\infty}^\infty\frac{d\lambda}{2\pi}\,i\lambda\,\mathrm{e}^{i\lambda(\frac{1} {z}-1-P_X\cdot m)}\,\delta(P_Y\!\cdot\! m)\,\langle 0|\,ig\,m_\eta F^{\eta \rho}(0)|Y\rangle\nonumber\\
&\hspace{1cm}\times\langle Y|\,q_i(0)|P_hS_h\,;\,X\rangle\langle P_hS_h\,;\,X|\bar{q}_j(0)\,|0\rangle\nonumber\\[0.3cm]
&\hspace{0.5cm}+\,\frac{1}{N_c}\sumintXY  \int_{-\infty}^\infty\frac{d\mu}{2\pi}\,(-i\mu)\,\delta(1/z-1+(P_Y-P_X)\!\cdot\! m)\,e^{-i\mu P_Y\cdot m}\nonumber \\
&\hspace{0.45cm}\times\langle 0|\,ig\,m_\eta F^{\eta \rho}(0)|Y\rangle\langle Y|\,q_i(0)|P_hS_h\,;\,X\rangle \langle P_hS_h\,;\,X|\bar{q}_j(0)\,|0\rangle\,. \label{DeltaFzz'Hc3}
\end{align}
As is discussed in detail after Eq.~(5) in~\cite{Meissner:2008yf}, the first term in Eq.~(\ref{DeltaFzz'Hc3}) is zero because $\delta(P_Y\!\cdot\! m)$ tells us that all the particles in the intermediate state $|Y\rangle$ (since they are on-shell) must be massless and move in the $n$-direction, and the matrix element 
$\langle 0|\,ig\,m_\eta F^{\eta \rho}(0)|Y\rangle$
vanishes in that case.  Therefore, we focus on the second term in Eq.~(\ref{DeltaFzz'Hc3}).  We note that $\sumintY \,\,\,= \sumintYone\,\,\,\cdots \sumintYN\,\,\,\,\,$ if there are $N$ particles in the intermediate state $|Y\rangle$.  We perform a change of variables so that one now has $\sumintY\,\,\,\to\sumintYp \,\,\,= \sumintYone\,\,\,\,\cdots \sumintYNonein\;\;\;\;\;\sumintY\,\,\,\,$.  This leads to 
\begin{align}
&\frac{\partial \Delta^{q,\rho}_{F,ij}(z,z^\prime)} {\partial (1/z^\prime)}\Bigg |_{z=z^\prime} \!\!=\frac{1}{N_c}\sumint\sumintYone\cdots \sumintYNone\,\sum_Y\int\!\frac{d(P_Y\!\cdot\! m)d^2\vec{P}_{YT}} {2P_Y\!\cdot\! m\,(2\pi)^3}  \nonumber\\
&\hspace{0.5cm}\times\int_{-\infty}^\infty\frac{d\mu}{2\pi}\delta(1/z-1+(P_Y-P_X)\!\cdot\! m)\,\frac{\partial} {\partial P_Y\!\cdot\! m}\left(e^{-i\mu P_Y\cdot m}\right)\nonumber \\
&\hspace{0.5cm}\times\langle 0|\,ig\,m_\eta F^{\eta \rho}(0)|Y\rangle\langle Y|\,q_i(0)|P_hS_h\,;\,X\rangle \langle P_hS_h\,;\,X|\bar{q}_j(0)\,|0\rangle \nonumber\\[0.3cm]
&= - \frac{1}{N_c}\sumint\sumintYone\cdots \sumintYNone\,\sum_Y\int\!\frac{d(P_Y\!\cdot\! m)d^2\vec{P}_{YT}} {2P_Y\!\cdot\! m\,(2\pi)^3} \,\nonumber\\
&\hspace{1cm}\times\delta(1/z-1+(P_Y-P_X)\!\cdot\! m)\,\delta(P_Y\!\cdot\! m)\nonumber \\
&\hspace{1cm}\times\langle 0|\,ig\,m_\eta F^{\eta \rho}(0)\frac{\partial} {\partial P_Y\!\cdot\! m}\bigg(|Y\rangle\langle Y|\bigg)\,q_i(0)|P_hS_h\,;\,X\rangle \nonumber\\
&\hspace{1cm}\times\langle P_hS_h\,;\,X|\bar{q}_j(0)\,|0\rangle\nonumber\\[0.3cm]
&\hspace{0.5cm} - \frac{1}{N_c}\sumint\sumintYone\cdots \sumintYNone\,\sum_Y\int\!\frac{d(P_Y\!\cdot\! m)d^2\vec{P}_{YT}} {2\,(2\pi)^3}  \,\nonumber\\
&\hspace{0.6cm}\times\frac{\partial} {\partial P_Y\!\cdot \!m}\bigg(\frac{1} {P_Y\!\cdot\! m}\delta(1/z-1+(P_Y-P_X)\!\cdot\! m)\bigg)\,\delta(P_Y\cdot m)\nonumber \\
&\hspace{0.6cm}\times\langle 0|\,ig\,m_\eta F^{\eta \rho}(0)|Y\rangle\langle Y|\,q_i(0)|P_hS_h\,;\,X\rangle \langle P_hS_h\,;\,X|\bar{q}_j(0)\,|0\rangle\nonumber\\[0.3cm]
&\hspace{0.3cm}+\, Boundary\; Terms\,, \label{DeltaFzz'Hc4}
\end{align}
where $Boundary\; Terms$ are from the partial integration over $P_Y\!\cdot \!m$.  The first and second terms in (\ref{DeltaFzz'Hc4}) vanish for the same reasons given above (after Eq.~(\ref{DeltaFzz'Hc3})), where the fact that $\partial/\partial(P_Y\!\cdot \! m)$ acts on $|Y\rangle$ in the first term does not alter the argument.  This leaves us only with the boundary terms to consider.  One has
\begin{align}
&Boundary\;terms = \frac{1}{N_c}\sumint\sumintYone\cdots \sumintYNone\,\sum_Y\int\!\!\frac{d^2\vec{P}_{YT}} {2P_Y\!\cdot\! m\,(2\pi)^3}  \nonumber\\
&\hspace{1cm}\times\int_{-\infty}^\infty\frac{d\lambda}{2\pi}\int_{-\infty}^\infty\frac{d\mu}{2\pi}\,e^{i\lambda(1/z-1+(P_Y-P_X)\cdot m)}\,e^{-i\mu P_Y\cdot m}\nonumber \\
&\hspace{1cm}\times\langle 0|\,ig\,m_\eta F^{\eta \rho}(0)|Y\rangle\langle Y|\,q_i(0)|P_hS_h\,;\,X\rangle\nonumber\\
&\hspace{1cm}\times \langle P_hS_h\,;\,X|\bar{q}_j(0)\,|0\rangle\bigg |_{P_Y\cdot m = (P_{Y_1} +\dots+P_{Y_{N-1}})\cdot m}^{P_Y\cdot m  = \infty} \label{DeltaFzz'Hc5}
\end{align}
The upper limit at $P_Y\cdot m = \infty$ in Eq.~(\ref{DeltaFzz'Hc5}) clearly vanishes.  The lower limit implies $P_{Y_N}\!\cdot m=0$ and also brings about a $\delta((P_{Y_1} +\dots+P_{Y_{N-1}})\cdot m)$ after the integration over $\mu$.  This means $P_Y\cdot m = 0$, and so again the matrix element $\langle 0|\,ig\,m_\eta F^{\eta \rho}(0)|Y\rangle$ will vanish.  Thus, we find $\frac{\partial}{\partial (1/z^\prime)}\Delta_{F}^{\rho}(z,z^\prime)\Big|_{z^\prime = z}=0$.  We note that higher derivatives of $\Delta_{F}^{\rho}(z,z^\prime)$ evaluated at $z^\prime = z$ will {\it not} vanish because the boundary terms in these cases are nonzero.


\end{document}